\documentclass[10, journal, compsoc]{IEEEtran}
\pdfoutput = 1
\usepackage{graphicx}
\usepackage{subfigure}
\usepackage{cite}
\usepackage{caption}
\usepackage{algorithm}  
\usepackage{algorithmicx}  
\usepackage{algpseudocode}  
\usepackage{amsmath}  
\usepackage{tikz}
\usepackage[hyphens]{url}
\usepackage{hyperref}

\begin{document}

\title{UrbanRhythm: Revealing Urban Dynamics Hidden in Mobility Data}
\author{Sirui~Song, 
        Tong Xia, 
        Depeng Jin~\IEEEmembership{Member,~IEEE,}
        Pan Hui~\IEEEmembership{Fellow,~IEEE,}
        Yong Li~\IEEEmembership{Senior Member,~IEEE,}
\IEEEcompsocitemizethanks{\IEEEcompsocthanksitem S. Song, T. Xia, D. Jin, Y. Li are with Beijing National Research Center for Information Science and Technology (BNRist) and with Department of Electronic Engineering, Tsinghua University, Beijing 100084, China.  E-mail: ssr15@mails.tsinghua.edu.cn, xia-t17@mails.tsinghua.edu.cn, \{jindp, liyong07\}@tsinghua.edu.cn.}
\IEEEcompsocitemizethanks{\IEEEcompsocthanksitem P. Hui is with the Department of Computer Science, University of Helsinki,00014 Helsinki, Finland, and also with the Department of Computer Science and Engineering, The Hong Kong University of Science and Technology, Hong Kong (e-mail: panhui@cse.ust.hk).}}

\IEEEtitleabstractindextext{
\begin{abstract}
Understanding urban dynamics, i.e., how the types and intensity of urban residents' activities in the city change along with time, is of urgent demand for building an efficient and livable city. Nonetheless, this is challenging due to the expanding urban population and the complicated spatial distribution of residents. In this paper, to reveal urban dynamics, we propose a novel system UrbanRhythm to reveal the urban dynamics hidden in human mobility data. UrbanRhythm addresses three questions: 1) What mobility feature should be used to present residents' high-dimensional activities in the city? 2) What are basic components of urban dynamics? 3) What are the long-term periodicity and short-term regularity of urban dynamics? 
In UrbanRhythm, we extract staying, leaving, arriving three attributes of mobility and use a image processing method Saak transform to calculate the mobility distribution feature. 
For the second question, several city states are identified by hierarchy clustering as the basic components of urban dynamics, such as sleeping states and working states. We further characterize the urban dynamics as the transform of city states along time axis. 
For the third question, we directly observe the long-term periodicity of urban dynamics from visualization. Then for the short-term regularity, we design a novel motif analysis method to discovery motifs as well as their hierarchy relationships.  
We evaluate our proposed system on two real-life datesets and validate the results according to App usage records. This study sheds light on urban dynamics hidden in human mobility and can further pave the way for more complicated mobility behavior modeling and deeper urban understanding.
\end{abstract}
}

\maketitle
\IEEEdisplaynontitleabstractindextext
\IEEEpeerreviewmaketitle

\section{Introduction}

As reported by UN\footnote{\url{https://www.un.org/development/desa/en/news/population/2018-revision-of-world-urbanization-prospects.html}}, up to 2018, $55\%$ of the world's population lives in urban areas, and this proportion is expected to increase to $68\%$ by 2050. 
The modern life style, the expanding urban population and the increasing complicated city structure bring changeable type, intensity and distribution of the residents activities, which which raise challenges to city management, ranging from traffic monitoring, resource scheduling to city planning. 

From the viewpoint of time, the changeable residents activities at different time lead the city transform between different states. For example, at rush hours when most residents are on the main road with crowded traffic, the city belongs to a state; while in working hours when most residents are concentrated in office area, the city belongs to another state. 
In order to build smart cities which are both efficient and livable, understanding how the transform of city state along with
the time, i.e., \textup{urban dynamics}, has become an urgent demand for policymaker, city governors and urban planners \cite{xia2019revealing}. 

Previous understanding of residents' activities comes from conducting surveys on human agents \cite{morenoff2001neighborhood}, which provides detailed information about people's behaviors.
However, collecting such kind of data is costly, and also has limitations in terms of generalization and geographical scope.
Luckily, smart phones and mobile network are popular and ubiquitous everywhere, which makes it available for us to collect large-scale mobility data. 
Recently, many works have investigated urban dynamics through resident' mobile behaviors. 
Sofiane \emph{et al.} \cite{spatiotemporaldynamics} built activity time series for London and Doha, and found that close neighborhoods tend to share similar rhythms. 
Louail \emph{et al.}  \cite{louail2014mobile} demonstrated that the city shape and hot-spots change with the course of the day. 
Fabio \emph{et al.} \cite{cityrhythm} captured the spatio-temporal activity in a city across multiple temporal resolutions, and visualized different activity levels in different time slots.
Xia \emph{et al.} \cite{xia2019revealing} revealed the daily activity patterns by learning offline mobility and online App usage together.  
However, these previous works are either based on statics \cite{spatiotemporaldynamics,louail2014mobile}, or case studies of several regions \cite{cityrhythm,xia2019revealing,fan2014cityspectrum}, which do not consider the spatial distribution of residents' activities in the city, thus are not able to present urban dynamics in a comprehensive and concise way.


To bring meaningful and useful insights in understanding urban dynamics, three key questions are raised:
\begin{enumerate}
\item What mobility features should be used to characterize urban dynamics from high-dimensional activities, considering that mobility has a spatial distribution in the city?
\item What are the basic components, i.e., city states, in urban dynamics?
\item What are the long-term periodicity and short-term regularity of urban dynamics?
\end{enumerate}

In this paper, we propose UrbanRhythm to address these questions. 

Firstly, after dividing the mobility data into different time slots, we look into the dynamics reflected by the mobility changing with these time slots.
Yuan et al. \cite{Zhenyu2015} has proved the moving-in and moving-out flow can be used to discover urban functional regions, and commuting is the most important activity in the city. Thus, for each region in the city in each time slot, we extract \emph{staying, leaving, arriving} three attributes to represent the mobility within it. 
Considering the spatial distribution of mobility in the city, for each time slot, we map the mobility of different regions in the city to a three-channel city image, where a pixel on the image represents a region, and three channels correspond to \emph{staying, leaving, arriving} attributes. 
Then image processing methods could be utilized to capture the mobility spatial distribution feature in the city. Compared with ordinary image processing tasks, we lack supervision and enough data to train a deep learning network. 
Thus we redefine an unsupervised image processing method Saak transform \cite{saak, lossysaak}, and utilize it to capture the mobility spatial distribution feature in the city.

To solve the second question, we detect city states, i.e., certain kinds of mobility distribution, by utilizing unsupervised clustering after calculating mobility distribution features for each time slot. 
Several city states are identified, such as working state, sleeping state, which highly correspond to residents' daily behaviors. 
The detected states could be further divided into sub-states. 
For example, sleeping state could be divided into deep-sleeping state and light-sleeping state.

For the third question, we first visualize the urban dynamics by full time mapping and 24-hour mapping to directly observe the long-term periodicity of urban dynamics. As a result, we find the long-term periodicity of urban dynamics highly correspond to the periodicity of weekdays, weekends and festival holidays. 
Then, to investigate the short-term regularity of urban dynamics, we design a novel motif analysis method and implement it on the city state series, discovering motifs of various lengths and their hierarchy relationships. We consider motifs in urban dynamics reflect residents' regular behaviors and  the composition pattern of motifs actually tells how residents' long-term regular behaviors are composed with short-term regular behaviors.

Finally, we carry out two experiments on two real-life datasets of Beijing and Shanghai. Besides, a validation experiment is done by employing a TF-IDF analysis \cite{Paik2013} on the relation between App usage and city states, which support our interpretation of the detected city states and further demonstrate that urban dynamics could be revealed from human mobility.

To summarize, the contribution of our work is four-fold:
\begin{itemize} 
    \item We propose a novel system UrbanRhythm to reveal daily urban dynamics. To the best of our knowledge, we are the first to consider the spatial distribution of mobility in the city when studying urban dynamics.
    \item We identify specific city states including working time, sleeping time, relaxing time, rush hours and other states corresponding to residents' daily life. These found states can be further divided into sub-states, like deep-sleeping and light-sleeping. 
    \item We find the long-term periodicity of urban dynamics correspond to weekdays, weekends and holidays. Besides, we use a novel motif analysis method to investigate the short-term regularity of urban dynamics, as well as their hierarchy relationships. This brings knowledge about residents' basic regular behaviors and how these behaviors compose long-term regular behaviors.
    \item We evaluate our method in two mobility datasets, Beijing and Shanghai, from different sources. A validation experiment is done by analyzing App usage records, which further support our revealing of urban dynamics. .
\end{itemize}


\section{Overview}


\subsection{Problem Statement}
In order to characterize urban dynamics from mobility data, we have the following definitions: 

\begin{figure}[ht]
\centering
      \subfigure[Grid-based city partition]{
        \includegraphics[width=0.18\textwidth]{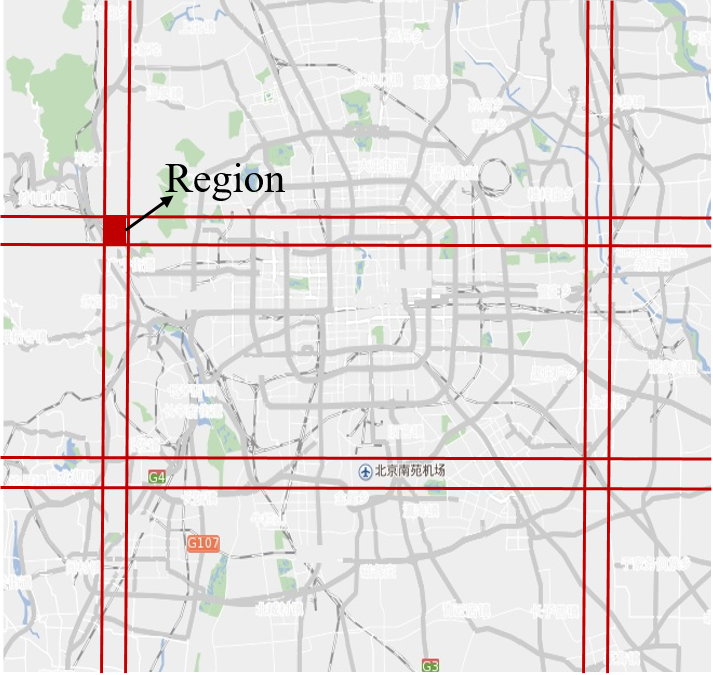}}\label{fig:definition1}
          \hspace{.2in}
      \subfigure[Three-channel city image]{
        \includegraphics[width=0.2\textwidth]{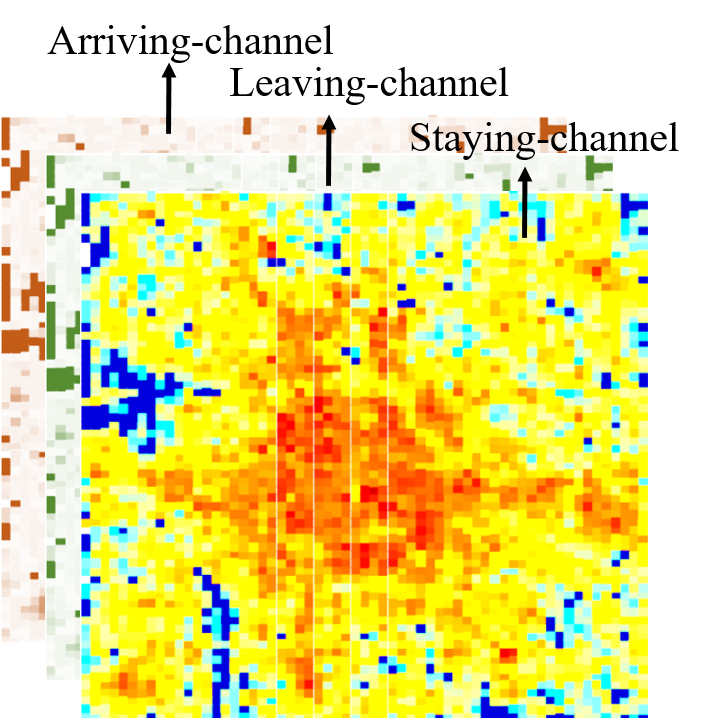}}
      \subfigure[City image series]{
        \includegraphics[width=0.48\textwidth]{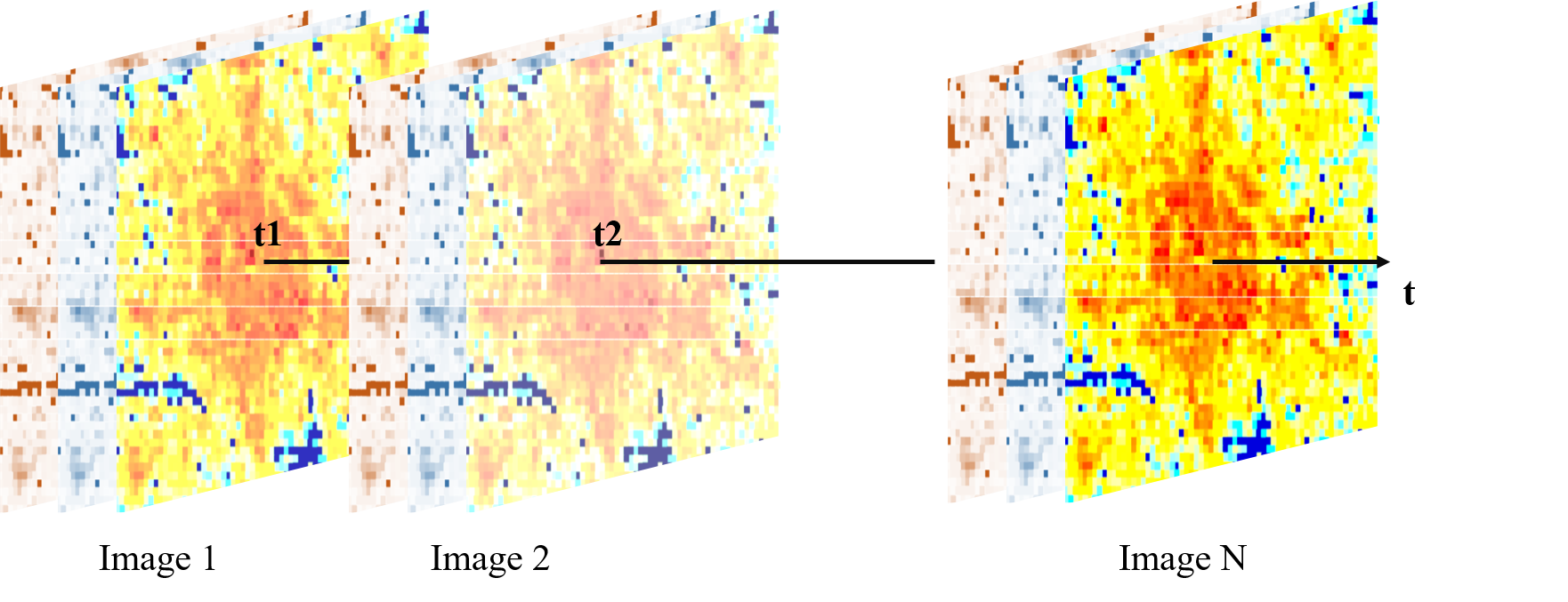}}
  \caption{A illustration of definitions of region, city image and city image series.}\label{fig:definition}
\end{figure}

\begin{figure}[t]
    \centering
    \includegraphics[width=0.42\textwidth]{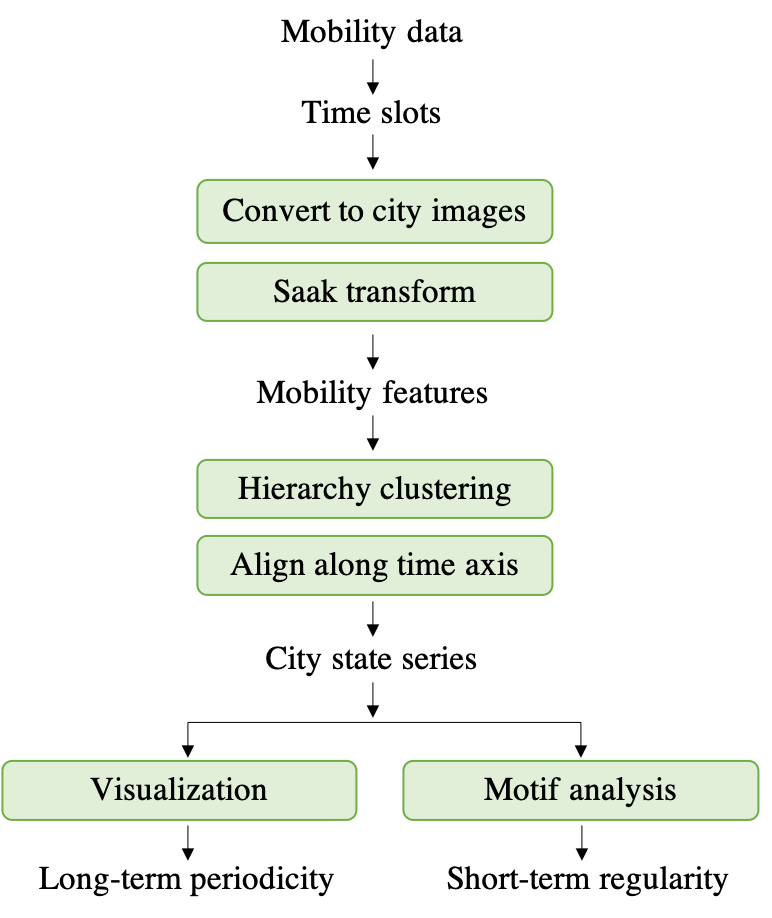}
    \caption{A illustration of UrbanRhythm system.}
    \label{fig:system}
    \vspace{-0.3cm}
\end{figure}

\textbf{Definition 1 (Region)} In this problem, we partition a city into a $X\times Y$ grid map based on the longitude and latitude where a grid denotes a region, as shown in Fig. \ref{fig:definition}(a). Here, region in $i$-$th$ row and $j$-$th$ column is denoted by $R_{i,j}$.

\textbf{Definition 2 (City Image, City Image Series)} After dividing a city into $X\times Y$ grids, we can describe the characters of the city by a three-channel image, where each channel presents one character and each pixel presents one region. Here we define the channels of an image as an staying-channel, a leaving-channel and an arriving-channel presenting that how many people stay at, leave from and arrive in the region during a given time slot, respectively \cite{Zhenyu2015}. The 3-channel image of a given time slot is shown in Fig. \ref{fig:definition}(b). 
City images at different time slots form a city image series, which reveal the variation of human mobility along with the time. A city image series is shown in Fig. \ref{fig:definition}(c), where $N$ is the total number of time slots. We denote the city image series by $I=\{I_1,I_2,...,I_n,..,I_N\}$ with $I_n$ denoting the image at $n$-$th$ time slot.

\textbf{Definition 3 (City State, City State Series)} We divide city images into several kinds. A city state represents a typical kind of city images and further represents a typical kind of mobility distribution. Similar city images share the same city state. We define the total number of city states to be $K$ and the state of city image $m_n$ to be $s_n$, where $s_n = 0,1,...,K-1$.

\textbf{Definition 4 (Urban Dynamics)} We classify each city image in city image series to a city state, forming a city state series $S = \{s_1,s_2,...,s_n,..,s_N\}$ with $s_n$ denoting the city state at $n$-$th$ time slot. We define urban dynamics as the transform of city states along with time.

\textbf{Definition 5 (Motif, Motif Class)} We define a sub-sequence of $S$ to be $S_{a,l}$, where $a,l$ denote the start time slot and the length of this sub-sequence, respectively. For $S_{a,l}$, if there exist one or more sub-sequences $S_{b,l}$ similar to $S_{a,l}$ and $b \neq a$, we call these sub-sequences motifs, and refer as $M_{a,l}, M_{b,l}$,...$M_{z,l}$ respectively. A motif class $C_n$ refers to a set of similar motifs of the same length. 

\textbf{Definition 6 (Motif relationship)} A motif could be a sub-sequence of another motif. We define the father/son relationships between motifs: 
If $M_{a_1,l_1}$ and $M_{a_2,l_2}$ are two motifs and $a_1 < a_2$ and $l_1 > l_2$, then call $M_{a_1,l_1}$ a father of $M_{a_2,l_2}$ and $M_{a_2,l_2}$ a son of $M_{a_1,l_1}$. For two motif class $C_i$ and $C_j$, if there exist $M_{a_1,l_1} \in C_i, M_{a_2,l_2} \in C_j$ and $M_{a_1,l_1}$ is the father of $M_{a_2,l_2}$, then we call  $C_i$ a father of $C_j$, and vice versa. In the following sections, without specification, we refer the relationships between motif classes as motif relationships.

In this paper, we aim to reveal urban dynamics. To do it, we answer three questions: what mobility feature to be used, what city states could be found, what long-term periodicity and short-term regularity would be. 

For the first question, we divide the mobility data into different time slots and aim to calculate the mobility feature within each time slot. 
For the second question, we aim to detect city states after extract mobility feature from each time slot and interpret the detected city states. 
For the third question, we aim to directly observe the long-term periodicity of urban dynamics from the visualization of the city state series and investigate the short-term regularity by analyzing motifs in the city state series.



\subsection{System Framework}


Our system is shown in Fig. \ref{fig:system}. 
To deal with the first question, we extract the mobility attributes \emph{staying, leaving, arriving} at different time slots to form the city image series $I$ and then conduct multi-channel Saak transform on each city image to calculate the spatial distribution pattern of mobility, get mobility features $V$.

To solve the second question, we employ hierarchical clustering on the mobility features $V$ to detect city states. Also, we interpret each city states according to the temporal distribution of states and the spatial distribution of the mobility.

After that, to answer the third question, we visualize urban dynamics by full time mapping and 24-hour mapping to observe long-term periodicity. Besides, we perform a novel motif analysis on the city state series to investigate the short-term regularity of urban dynamics, which includes three parts: discovering motifs, determining motif classes based on the discovered motifs, and finally investigate the hierarchical relationship of motif classes. 

In the end, we take an App usage analysis to validate our detection and interpretation of city states.

\section{Algorithm Design}
In this section, we introduce the algorithms used in this paper, including calculating mobility features by Saak transform, extracting city states by hierarchy clustering, investigating the long-term periodicity and short-term regularity of urban dynamics by visualization and motif analysis.



\subsection{Calculate mobility features}

\begin{figure*}[ht]
\centering{\includegraphics[width=0.9\textwidth]{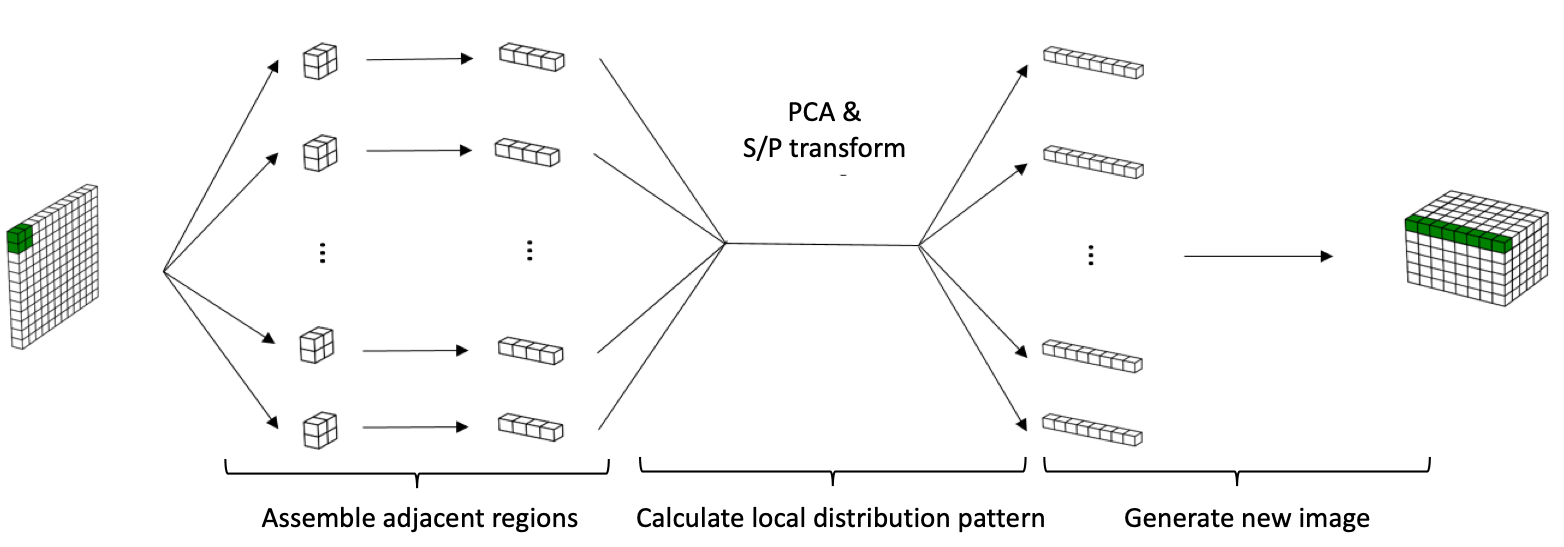}}
\caption{The first stage of Saak transform. 
We assemble each four adjacent regions into a grid, then apply PCA on all grid vectors and conduct a S/P transform on the outputs vectors of PCA. Finally, we refill the transformed vectors to their original grids and generate new images.
}
\vspace{-0.2cm}
\label{fig2}
\end{figure*}

To calculate the spatial distribution of mobility in the city, our basic method is Saak transform. In this section, we introduce Saak transform under our problem definition and how we adapt it to our multi-channel city images.

Kuo and Chen \cite{saak} proposed Saak transform recently. Saak transform converts a single-channel image $A_{n}$ to a feature vector $V_{n}$ in spectral space by implement Karhunen-Loeve transform (KLT) step by step. Chen et al. \cite{lossysaak} put forward lossy saak transform, in which he uses the principal component analysis (PCA) instead of KLT to save time and space.


Under our problem definition, with images series as input, each stage of Saak transform has the following three steps:

1) \textbf{Assemble adjacent regions:} We first choose the size of area in which we calculate the local distribution pattern. In practice, we choose the basic scale of $2 \times 2$. Let value in region $R_{i,j}$ denoted by $r_{i,j} \in R^D$, $i,j = 1,2\dots L_{in}$, where $L_{in}$ is the input width and height. For each city image, assemble each 4 adjacent regions to be a new grid, denoted as $G_{i,j} \in R^{4D}$, $i,j = 1,2 \dots L_{out}$.
\begin{gather}
    g_{i,j} = Concatenate(r_{2i-1,2j-1},r_{2i-1,2j},r_{2i,2j-1},r_{2i,2j}) \\
    L_{out} = L_{in}/2
\end{gather}

2) \textbf{Calculate local distribution pattern:} We conduct principal component analysis (PCA) on grids vectors from all $N$ city images. In this way, for each grid $G_{i,j}$, a comparison with other grids among all city images is implemented and the variation pattern is calculated and expressed as output vectors $G^{*}_{i,j}$. 

To avoid the change of sign in two consecutive stage, we conduct a sign-to-position(S/P) transform , with $G^{*}_{i,j}$ as input and $G'_{i,j}$ as output. 
\begin{align*}
    g'_{2k-1} = ReLU(g^{*}_{k}), k=1,2 \dots 4D \\
    g'_{2k} = ReLU(-g^{*}_{k}), k=1,2 \dots 4D
\end{align*}

3) \textbf{Generate new image:} Refill each gird $G_{i,j}$ with the transform vector $g'_{i,j} \in R^{8D}$. Form $N$ new images with half the original width and height. The spatial relationship between grids are kept for the next stage transform.

The scale of $2 \times 2$ is the smallest scale we can choose. Using bigger scale like 3*3 or 4*4 may miss the influence of small district pattern to city state. And the same as \cite{lossysaak} did, we reserve components with explained variance ratios lager than $3\%$ in PCA, which has been proved to be an acceptable compromise between efficiency and reserving discriminative information \cite{lossysaak}. 

The first stage of Saak transforms is illustrated in Fig. \ref{fig2}. In $k$ stage of Saak transform, the local spatial pattern of $2^k \times 2^k$ scale is calculated. Put together the outputs of all stages, the spatial distribution pattern of mobility is calculated.


\textbf{Multi-channel Saak transform} The original Saak transform only deals with one channel at one single time. We can't directly concatenate three channels of city images and apply Saak transform because people's staying, leaving, arriving are obviously correlated. 
Thus we design Multi-channel Saak transform. We apply KLT on channels to do decorrelation and use KLT-transformed images as input for Saak transform. For each city image, put together the outputs for all stages of Saak transform as the feature vector for this city image.



\subsection{Extract city states}

After Saak transform, each city image $I_{n}$ can be represented as feature vectors $V_{n}$($n=1,2\dots N$). 
To save time and space for clustering, we apply PCA on feature vectors to reduce their dimensions to 128, uniformly. The choice of this dimension is under the consideration of the explained variance ratio of PCA.

Intuitively, human mobility behaviors usually have intrinsic periods of day and week; city state of different time could be alike. Thus unsupervised clustering method can be utilized on city images to find those with similar mobility features.
However, the problem of totally unsupervised clustering is that we don't have a specific standard to evaluate the cluster results and due to that it's hard for us to specify a number of clusters. On the other hand, we're not only curious about a specific set of city states or a specific kind of city dynamics, but also their inclusion relationships. So to better understand the process of clustering and the relationship between clusters, we use hierarchical clustering method to cluster feature vectors. 

We conduct hierarchical clustering in these obtained feature vectors $V_{n}$($n=1,2\dots N$) of city images. The basic idea of hierarchical clustering is to generate a tree of clusters where two son clusters merge to form a father cluster. The leaf node of this tree is the input $N$ feature vectors. And then from bottom to up iteratively merge the most suitable two clusters until the stop condition is met. 
We define the suitability of two clusters' merging according to Ward's method \cite{Jr1963Hierarchical}, to minimize the variance of the clusters after merging. By applying hierarchical clustering instead of distance-based or density-based clustering, we could analyze the dynamic states at different levels.

\subsection{Visualization}

To directly observe the periodicity of urban dynamics, we visualize the obtained urban dynamics in two aspects as follows:

1) \textbf{Full time mapping:} 
We plot the obtained state series along time axis, presenting the transform of city state over time. By doing this, we hope to reveal the period of urban dynamics and roughly locate the times when the city experiences different dynamics.

2) \textbf{24-hour mapping:} 
We show each 48 time slots in the same day as a 24-hour pie chart. Besides, according to the coarse localization of different dynamics, we divide the time slots into weekend, weekday and holidays to show 24-hour pie charts respectively. By doing this, dynamics within a day can be observed and different kinds of dynamics are presented and compared.

\subsection{Motif analysis}


To investigate the short-term regularity of urban dynamics in a fine-grained style, we design a motif analysis method, which contains three parts: discovering motifs, determining motif classes, and investigating motif relationships. Among the three parts, the first part is inspired by \cite{Motif}. 

\textbf{Discovering motifs:} To discovery motifs of arbitrary length, we use a divide-merge strategy. A pseudocode of discovering motifs is shown in Algorithm 1, which includes the following steps:

1) Cut the city state series into windows of length $l_w$, with stride $s_w$. Denote the sub-sequence within the $i-th$ window as $w_i$.

2) Compare the similarity between windows and map the results into a collision matrix $Mat$, where
\begin{equation}
    \begin{aligned}
        Mat(i, j)=\left\{
        \begin{array}{rcl}
        True && {Dis(w_i, w_j) \leq \sigma_w}\\
        False && {Dis(w_i, w_j) > \sigma_w}
        \end{array}\right. \\
    \end{aligned}
\end{equation}
$Dis$ is the Hamming distance function. 

3) Extract traces from $Mat$. Each trace correspond to two similar sub-sequences, which are composed with several windows. We transfer traces to motifs $M_{i,l}$.

4) Transfer the extracted traces to motifs $M_{i,l}$.

\begin{algorithm}
\label{alg:1}
\caption{Discovering motifs from city state series} 
\begin{algorithmic}
\Require City state series $S$, window length $l_w$, window difference threshold $\sigma_w$, stride $s_w$.
\Ensure Motifs $M$. \\
\\
1) Cut $S$ into windows:
\For{$i \in range(0,len(S),l_w)$} 
    \State $ w[i] = S[i*s_w:i*s_w + l_w] $ 
\EndFor \\

\\
2) Compare windows and map the results into $Mat$:
\For{$i \in range(0,len(w))$} 
    \For{$j \in range(i,len(w))$}  
        \State $ Mat[i,j] = Similar(w_i,w_j)$
    \EndFor
\EndFor \\

\\
3) Extract traces from $Mat$:
\State $traces = \{\}$
\For{$i $ in $ range(len(Mat))$}  \verb|\\|Find traces
    \For{$i $ in $ range(len(Mat))$}
        \If{$Mat[i,j] == True$}
            \State $L = 1$
            \While{$Mat[i+1,j+1] == True$}
                \State $L = L+1$
            \EndWhile
            \State $trace[L].append(i)$
        \EndIf
    \EndFor
\EndFor \\

\\
4) Convert traces to motifs:
\State $M = \{\}$
\For{$L$ in $traces$}   \verb|\\|convert traces to motifs
    \For{$start$ in $traces[L]$}
        \State $i = start * s_w$
        \State $l = l_w + (L-1) * s_w$
        \State $M[l].append(S[i:i+l])$
    \EndFor
\EndFor \\
\State \Return{$M$}
\end{algorithmic}
\label{alg}
\end{algorithm}

\textbf{Determine motif classes:} We use DBSCAN clustering \cite{ester1996density} to cluster motifs of same length. Define each cluster to be a motif class $C_n$.

\textbf{Investigate motif relationships:} We investigate the inclusion relationships of motifs classes and visualize them with the following steps:

1) For each pair of motif classes, determine their relationship according to Definition 6.

2) For each motif class, delete its “grandson”: if $C_x$ is a father of $C_y$, and both of them are sons of $C_z$, then delete $C_y$ from sons of $C_z$.

3) Build a directed graph based on motif relationships.

By employing the above analysis method, motifs of different lengths will first be discovered and then be categorised into different classes. Finally, a directed graph is build to present the hierarchy relationship of motifs.






\section{Performance Evaluation}
In this section, we evaluate our algorithm in two different datasets and investigate the basic components and the regularity in urban dynamics.

\begin{table*}[ht]
\begin{center}
\begin{tabular}{c|cccc}
\hline
\textbf{City}&   \textbf{Sources} &   \textbf{Localization Method} & \textbf{Duration} & \textbf{Number of Users} \\
\hline
 \textbf{Beijing, China}  & Mobile applications &  GPS module & 1 Apr.$\sim$30 Apr. (2018) & 18,916,166  \\
  \textbf{Shanghai, China} &  Cellular network &  Cellular base station & 21 Apr.$\sim$25 Apr. (2016) & 1,700,000 \\
  \hline
\end{tabular}
\end{center}
\vspace{-0.2cm}
\caption{Key features of two real world mobility datasets we utilize.}\label{tab:data}
\vspace{-0.4cm}
\end{table*}

\subsection{Datasets}
We collect two large scale real world mobility datasets to apply and evaluate our methodology. The datasets are collected from two different metropolis: Beijing and Shanghai, China. The features of the datasets are presented in Table~\ref{tab:data}. Shanghai dataset also contains the mobile applications (App) the mobile users are currently using, by resolving the $URI$ of $HTTP\ requests$. We use this App usage records to further validate the city states identified by analyzing mobility features.



\textbf{Beijing:} This dataset is collected from the mobile devices in Beijing by a popular mobile application vendor. It records the spatio-temporal information of mobile users whenever they request localization services in the applications, such as check-in and location-based social network. The localization of the mobility records is mainly achieved by GPS modules on the mobile devices plus network-based enhancement. 
This dataset is large scale in terms of tracing 18,916,166 mobile users in one month. 

\textbf{Shanghai:} This dataset is collected by a major cellular network operator in Shanghai, China. 
It is a large scale mobility dataset also covering 1,700,000 mobile users with the duration of 5 days. 
It records the spatio-temporal information of mobile subscribers when they access cellular network.
(i.e., making phone calls, sending texts, or consuming data plan). 
Thus, the recorded locations are at the granularity of cellular base stations. 
It also contains the mobile applications (App) the mobile users are currently using, by resolving the $URI$ of $HTTP\ requests$. 
Such associations also provide insights about understanding the dynamics from another angle. There are 10 types of Apps in our dataset: Social, Video, Music, Reading, Game, Shopping, Restaurant, Transportation, Office, Stock.

\textbf{Privacy and ethical concerns:} We have taken the following procedures to address the privacy and ethical concerns of dealing with such sensitive data. First, all of the researchers have been authorized by the application vendor and cellular network operator to utilize these two datasets for research purposes, and are bounded by strict non disclosure agreements. Second, the data is completely anonymized by replacing the users' identifiers with random sequence. Third, we store all the data in a secure off-line server, and only the core researchers can access the data.

\subsection{Pre-processing}

We divide Beijing into $1km\times km$ grids, and finally remain the areas in downtown with total grid number of $61\times 65$. 
For Shanghai, to evaluate the flexibility of our framework, we divide its city areas into $256 \times 256$ grid map, where each grid has a granularity of $200m\times 200m$. 
Besides, we calculate the mobility features \emph{staying, leaving, arriving} for each half hour. Thus, the number of city images for Beijing is 1440 and for Shanghai is 240. 

\subsection{Feature Space Visualization}
\begin{figure}[t]
  \centering
        \subfigure[t-SNE for Beijing]{
        \includegraphics[width=0.21\textwidth]{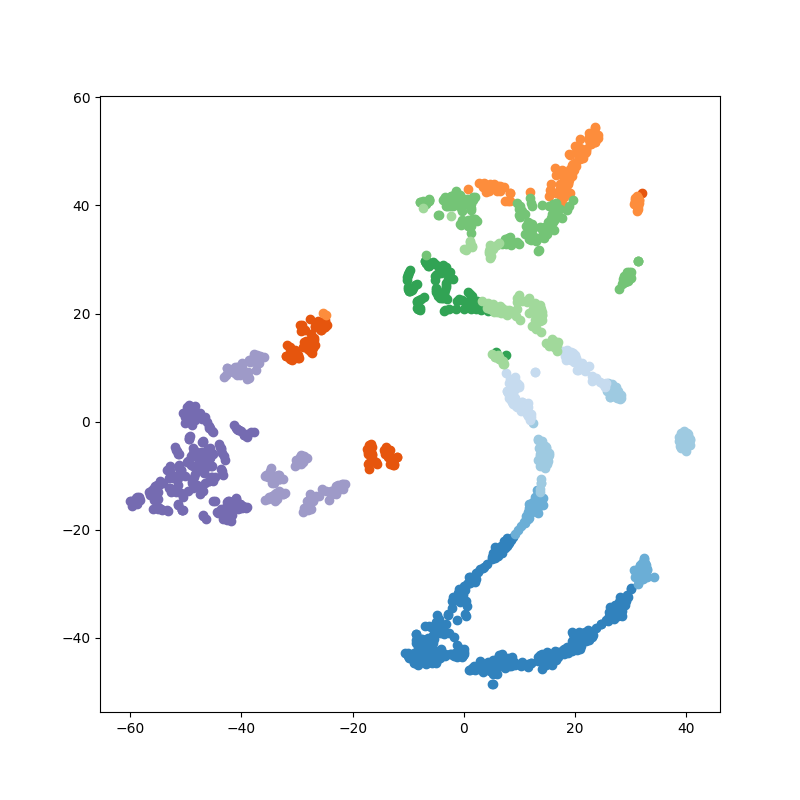}}
      \subfigure[t-SNE for Shanghai]{
        \includegraphics[width=0.21\textwidth]{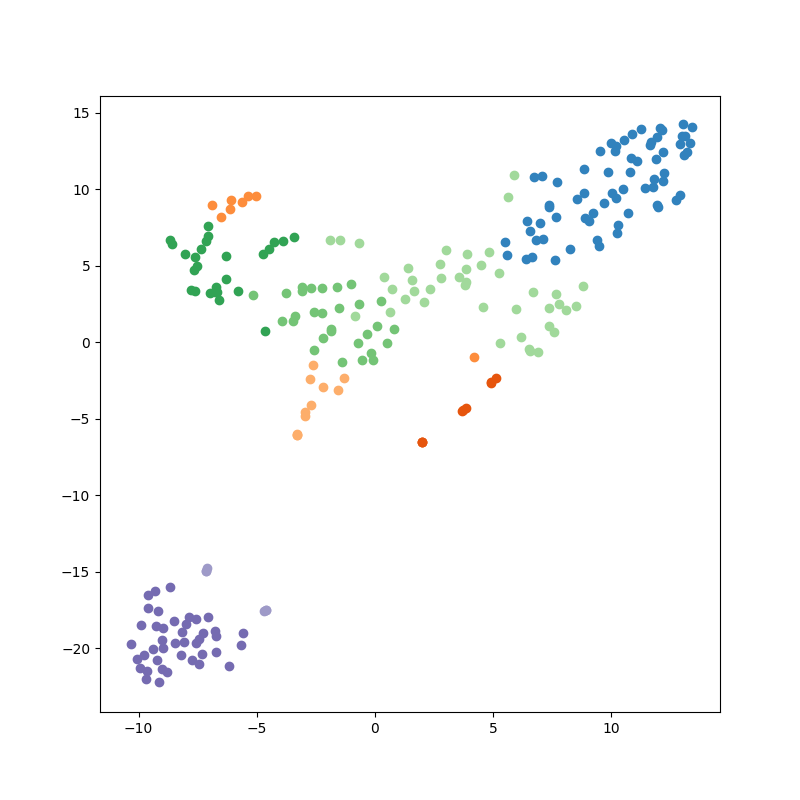}}
  \caption{Feature visualization using 8 clusters to represent 8 states, where each point represents one time slot and time slots in the same state are presented in the same color.}\label{fig:features}
\vspace{-0.2cm}
\end{figure}

We apply PCA on features vectors after Saak transform to reduce their dimensions and as the input for clustering. 
We conduct t-SNE \cite{tsne} to visualize the relationship of all 128-dimensional features. Results for Beijing and Shanghai are shown in \ref{fig:features}(a) and \ref{fig:features}(b), respectively.
From these two figures, we can explicitly observe that in the feature space, the time slots of the same state distribute closely to each other, while the time slots of different states generally have a larger distance. Therefore, it demonstrates that the Saak and PCA transform is effective to represent the feature of time slots.

\begin{figure}[ht]
  \centering
      \subfigure[Beijing clustering]{
        \includegraphics[width=0.42\textwidth]{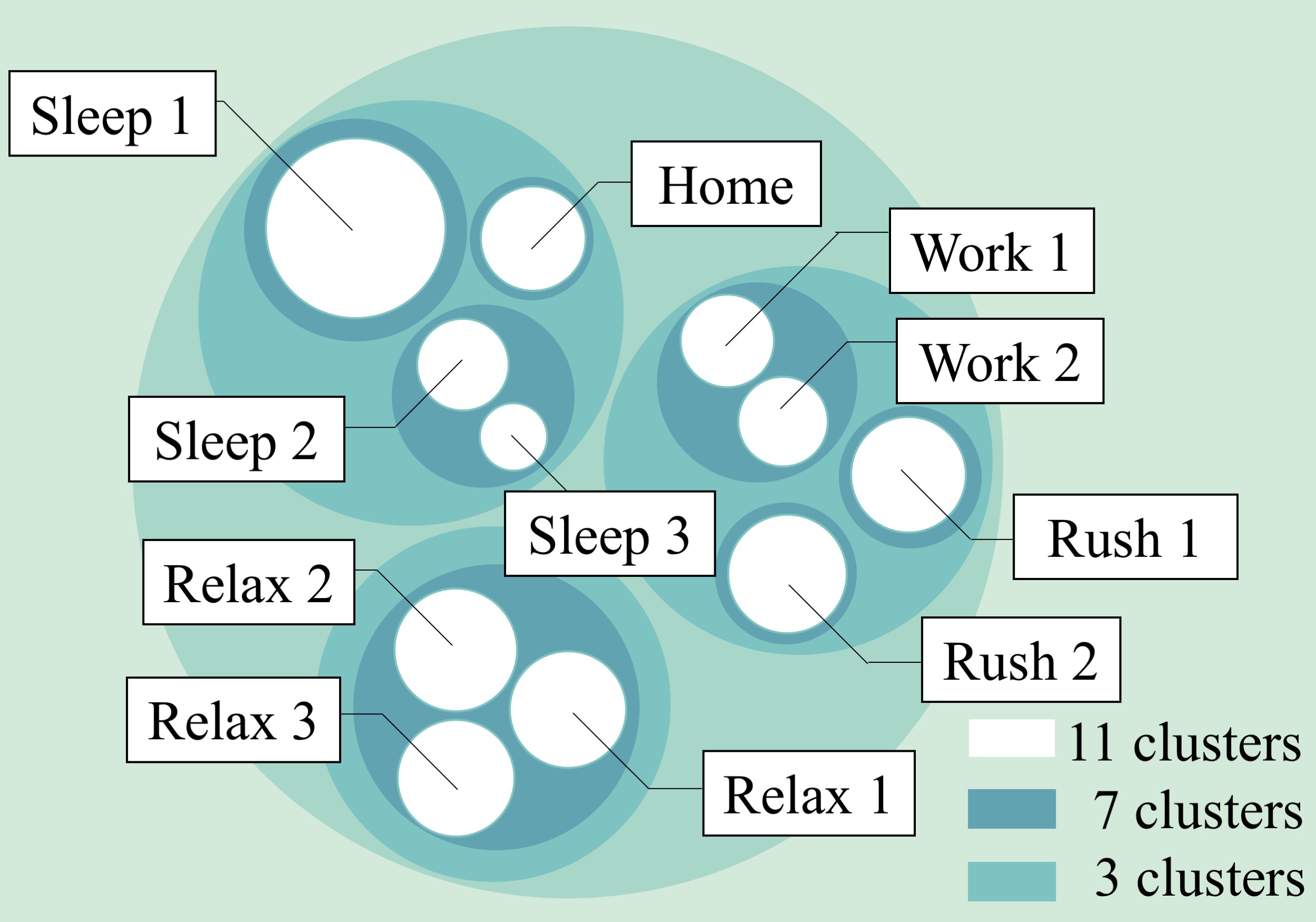}}
        \hspace{.3in}
      \subfigure[Shanghai clustering]{
        \includegraphics[width=0.42\textwidth]{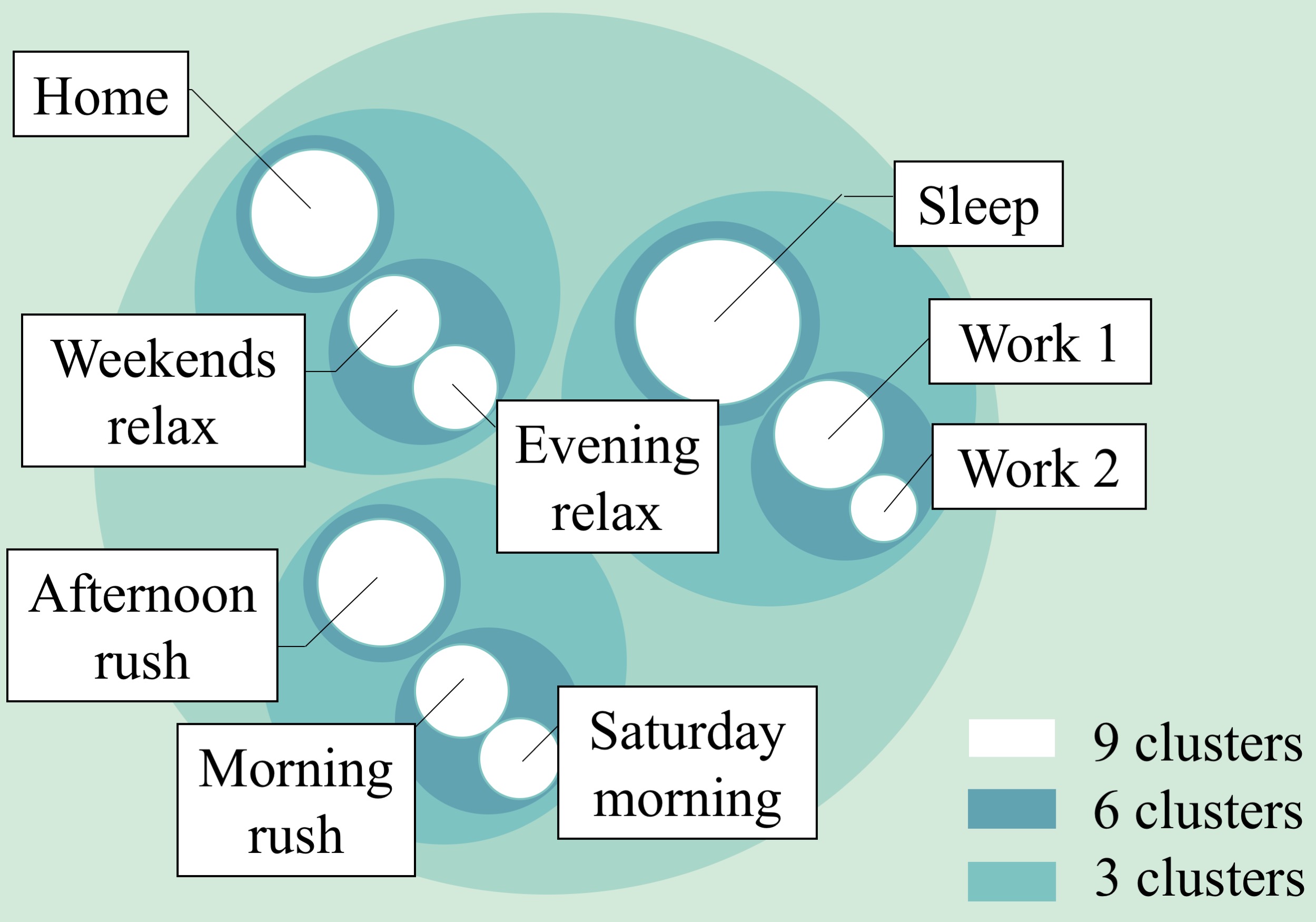}}
  \caption{Hierarchical clustering results with different cluster numbers.} 
  \label{fig_circle}
\vspace{-0.2cm}
\end{figure}

\subsection{Revealing urban dynamics}

\subsubsection{Hierarchical Clustering Structure}
Since hierarchical clustering is utilized, the structure of clustering results from up to bottom could be clearly observed.
By default, we display the cluster hierarchy using several circles, where child clusters are nested within their parent cluster. 
This gives a clear view of the hierarchical relationships of different clusters. Circle sizes reflect the number of time slots in the cluster, which allows us to quickly identify the most prevalent states. 

For Beijing, we show the 3-level results for 3, 7, 11 clusters exhibited in circles with the color from blue to white in Fig. \ref{fig_circle}. We also label the semantics for each state when the time slots are divided into 11 clusters. Obviously,  the outermost three circles represent three basic states in city that people are working, relaxing and sleeping. When the number of clusters increases, the time slots can be divided into more detailed states. For example, the basic sleeping state of Beijing can be divided into four states \emph{Home}, \emph{Sleep 1}, \emph{Sleep 2} and \emph{Sleep 3}, which represent different levels of people's staying home and movement in the city. The latter three sub-states could be further interpreted as different levels of how many people are sleeping, respectively. The same is to Shanghai. We show the 3-level results for 3, 6, 9 clusters in Fig. \ref{fig_circle}.

To conclude, the hierarchical relationships of different time slots is consistent with our intuitions to the states of city, which is pave the way for our understanding of urban dynamics.

\begin{figure}[ht]
  \centering
    \subfigure[Beijing state series]
        {\includegraphics[width=0.45\textwidth]{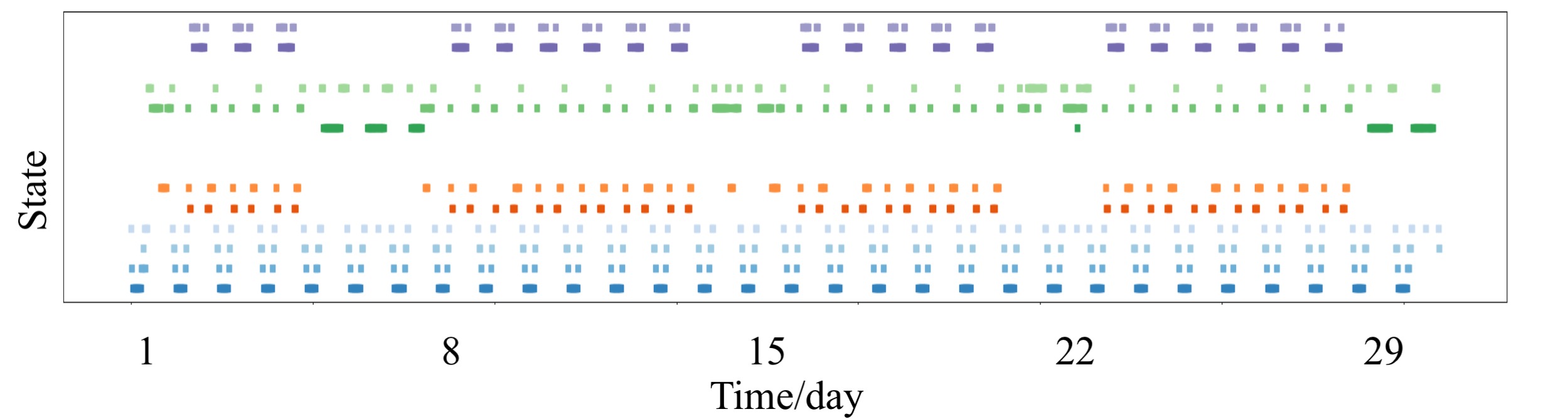}}
      \subfigure[Beijing 24 hours pie charts]
        {\includegraphics[width=0.49\textwidth]{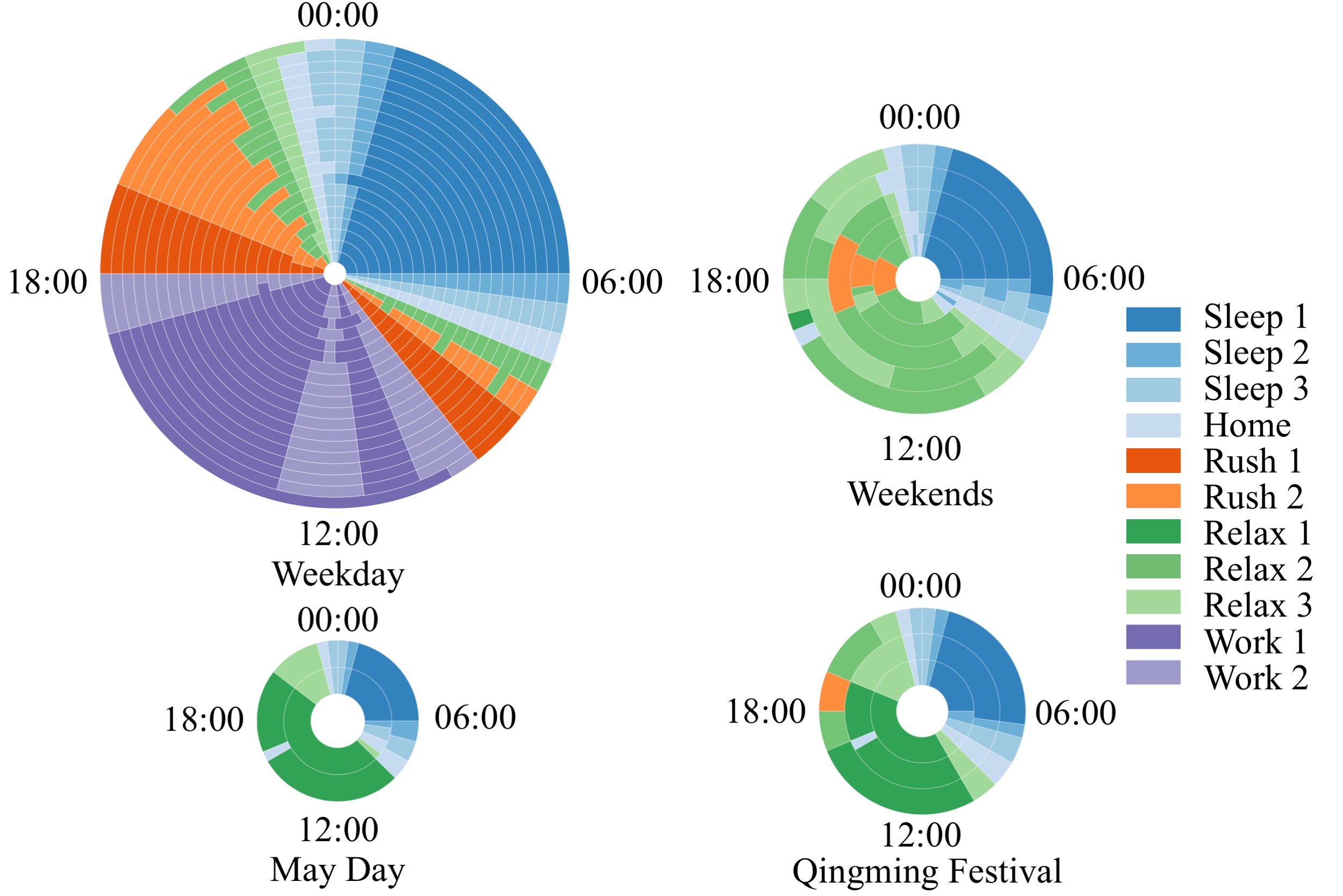}}
  \caption{Visualization of dynamics for Beijing when the number of clusters is 11. In (a), we show the transform of city state along with time within 30 days. In (b), we visualize dynamics for 4 kinds of days, i.e., weekday, weekends, May Day and Qingming Festival. }  \label{fig:bj_clustering}
\end{figure}

\begin{figure}[ht]
  \centering
    \subfigure[Shanghai state series]
        {\includegraphics[width=0.45\textwidth]{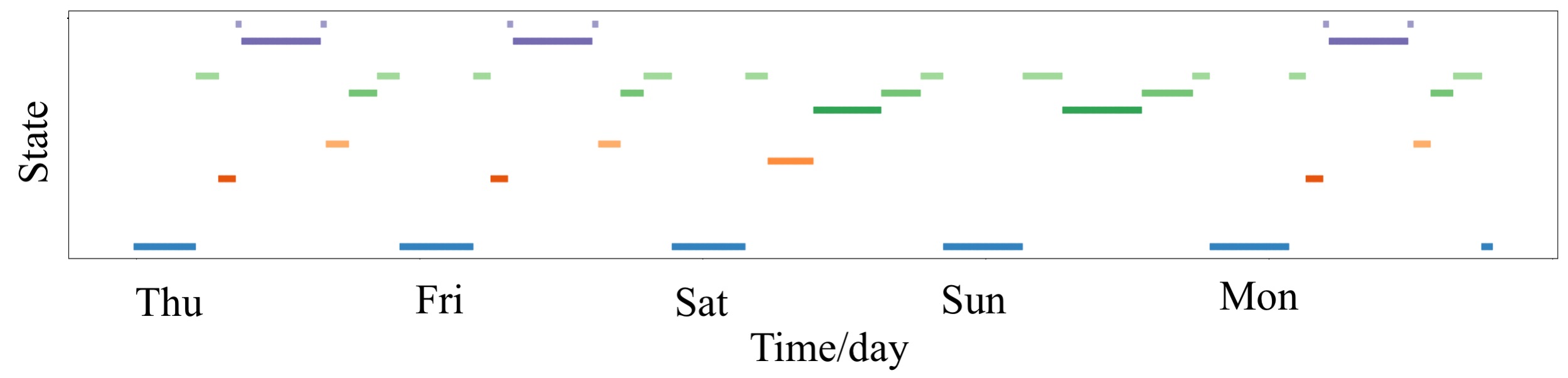}}
      \subfigure[Shanghai 24 hours pie charts]
        {\includegraphics[width=0.49\textwidth]{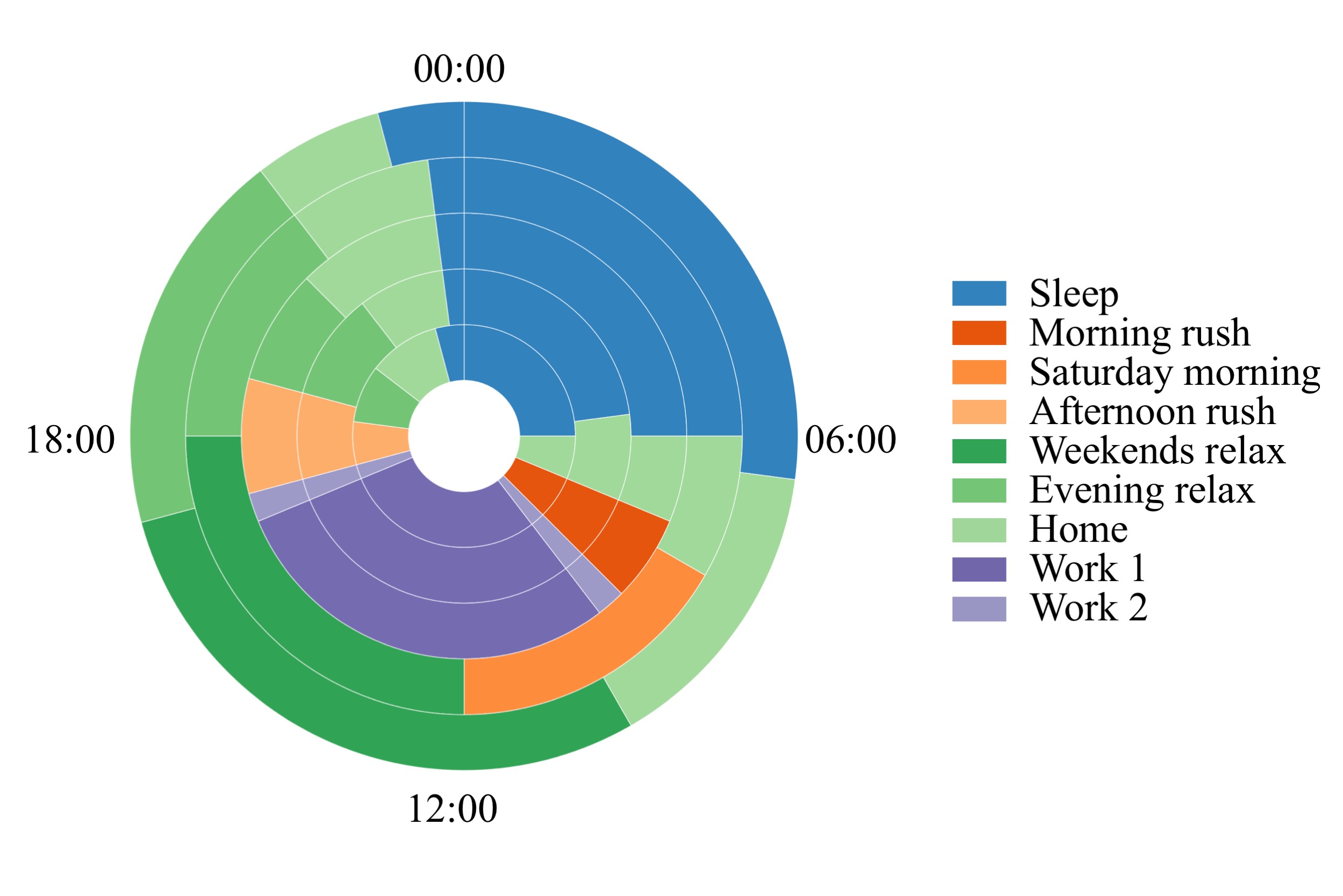}}
  \caption{Visualization of dynamics for Shanghai when the number of clusters is 9. In (a), we show the transform of city state along with time within 5 days. In (b), to better compare dynamics between weekdays and weekends, we visualize dynamics for Mon, Thu, Fri, Sat, Sun from inner circle to outer circle.}  \label{fig:sh_clustering}
\vspace{-0.2cm}
\end{figure}

\subsubsection{The extracted city states}

To analyze specific city states and investigate how they correspond to residents' daily life, we set the number of clusters to be 11 and 9 for Beijing and Shanghai respectively and interpret the physical meaning of each state by analyzing the temporal distribution pattern of states, the spatial distribution pattern of mobility, and the relationship between states and sub-states. We also show the full time mapping and 24-hour mapping visualization of city state series in Fig. \ref{fig:bj_clustering} and Fig. \ref{fig:sh_clustering}.




For Beijing, the state series is shown in Fig. \ref{fig:bj_clustering}(a), and the 24-hour pie chart is shown in Fig. \ref{fig:bj_clustering}(b). Since the dataset of Beijing covers a whole month, we can easily observe the period of day and week in the state transform process.  The distribution of state on the time axis is very symmetrical and neat, which is consistent with the regularity of people's daily commuting. To explain these states in more detail, we align the states on the time axis and display them in 24-hour pie chart, where each circle presents one day and time slots in the same state are exhibited in the same color. We summarize the characters of each state as follows:

\textbf{Sleep States}: These states include Sleep 1, Sleep 2 and Sleep 3. 
In these states, most people are sleeping and few people are moving in the city, reflected by bigger value in staying-channel than arriving-channel and leaving-channel. Besides, values in all three channels in Sleep states are much smaller than others states, suggesting few people are using the mobile application. Values in all three channels decrease from Sleep 3 to Sleep 2 to Sleep 1, which means more and more people become asleep.

\textbf{Home State}: This state usually covers 23:00-23:30 and 7:00-7:30 in all days. It is similar to Sleep states with larger value in staying-channel and smaller value in leaving-channel and arriving-channel, according to the clustering structure in Fig. \ref{fig:bj_clustering}. However we are surprised to find it also appears in some non-weekdays afternoons. 

\textbf{Rush States}: These states include Rush 1 and Rush 2.
In these states, most people are moving in traffics, reflected by bigger value in arriving-channel and leaving-channel than staying-channel. The distribution of people in city address the main road. Specifically, Rush 1 only appears in weekdays, corresponding to go-to-work and off-work rush. Rush 2 appears both weekdays are non-weekdays. Compared to Rush 1, people presents more staying, less leaving and arriving.

\textbf{Work States}: 
These state include Work 1 and Work 2 state, both appearing in only weekdays.
In these state, most people are working reflected by high values of official areas in staying-channel. Thus, we conclude in these states most people are at work. Besides, in Work 2, people's movement is more frequent than in Work 1. We are surprised to find that people's movement in noon is close to that in the beginning and end of office time.

\textbf{Relax States}: These states include Relax 1,2,3.
Relax 1 covers most day-time in holidays when many people travel far away from the city. Relax 2 covers day-time in weekends, 22:00-22:30 and 7:30-8:00 in weekdays and it presents larger value in all three channels than Relax 1. Relax 3 appears mostly after Relax 2 or in non-weekdays mornings, with much lower arriving value and leaving value than Relax 2. \\

For Shanghai, the state time series is shown in Fig. \ref{fig:sh_clustering}(a), and the 24-hour pie chart is shown in Fig. \ref{fig:sh_clustering}(b). Since the dataset of Shanghai covers only five days, we can only observe the period of day.  But the distribution of state on the time axis is still very symmetrical and neat. We summarize the characters of each state as follows:

\textbf{Sleep State}: This state mainly covers 23:30-06:00. Most people are sleeping and few people are moving in the city. Values in arriving-channel and leaving-channel are very low.

\textbf{Work States}: These states include Work 1 and Work 2. Most people are at work with slight movement in the specific office district. Specifically, people in Work 2 state have more movement than Work 1.

\textbf{Rush States}: These states include Morning rush, Afternoon rush and Saturday morning. In these rush states, people's moving is much stronger than work and sleep states. Movement in theses states addresses downtown areas. The value of leaving-channel in Morning rush higher than that of arriving-channel. However, it is just opposite in Afternoon rush. In Saturday morning, value in both leaving-channel and arriving-channel is high, suggesting the movement in Saturday morning is more directionless than that in weekdays.

\textbf{Relax States}: These states include Weekends relax and Evening relax. The movement is more frequent than work hours and less frequent than rush hours, as well as less concentrated in office areas and downtown areas. This indicates people are moving all around the city without very heavy traffic. Thus we conclude people are moving for relaxing in these two states. 

\textbf{Home State}: This state usually covers 21:30-23:30 and 6:00-7:00 in all days. It is similar to Relax states, for they belong to the same root state according to Fig. \ref{fig_circle}. However the values of leaving-channel and arriving-channel are smaller than that in Relax states, but larger than that in Sleep states. Thus Home state corresponds to the time when people are at home with few movement. 


To conclude, observing the state in the 24-hour pie chart from clockwise, we have that the dynamics of city from morning to night, from day to month, which reveal the regularity of people's mobility behavior from inactive to active, and last back to inactive in circle of one day.



\begin{figure}[ht]
\centering
    \subfigure[Go-to-work rush]{
        \includegraphics[width=0.46\textwidth]{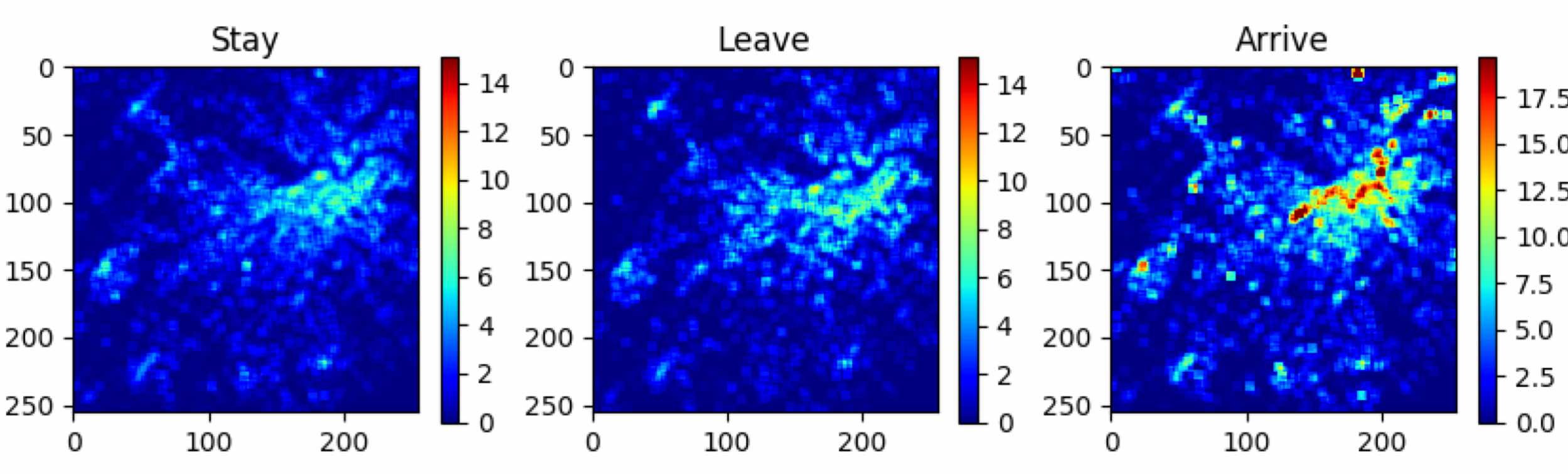}}\label{fig:sh_go_to_work}
    \subfigure[Off-work rush]{
        \includegraphics[width=0.46\textwidth]{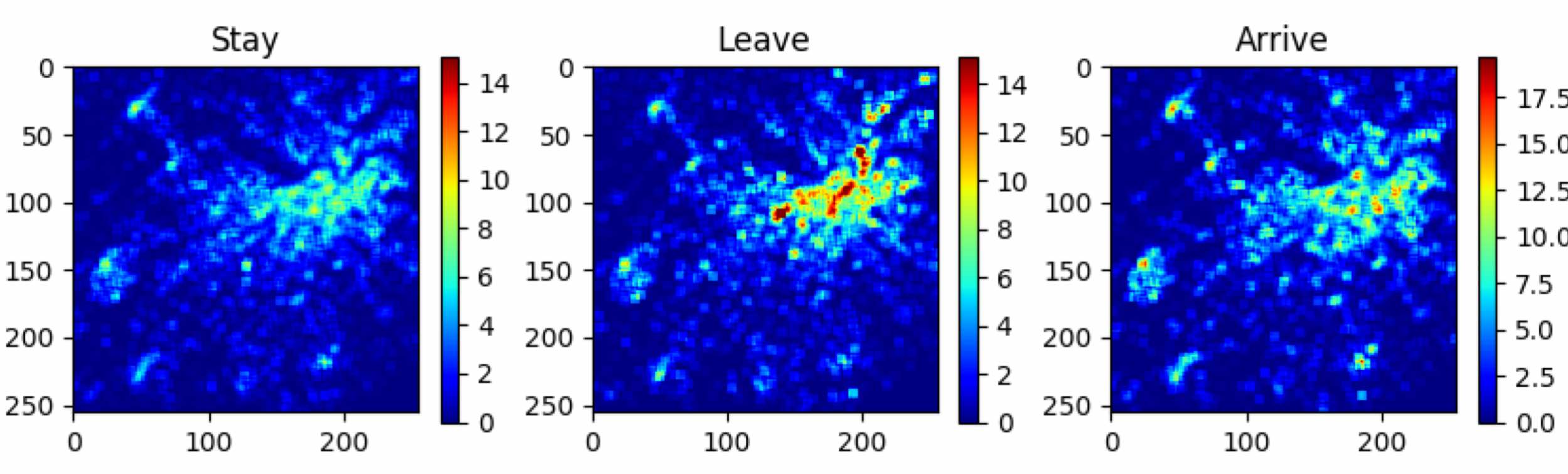}}\label{fig:sh_off_work}
    \subfigure[Sleeping]{
        \includegraphics[width=0.46\textwidth]{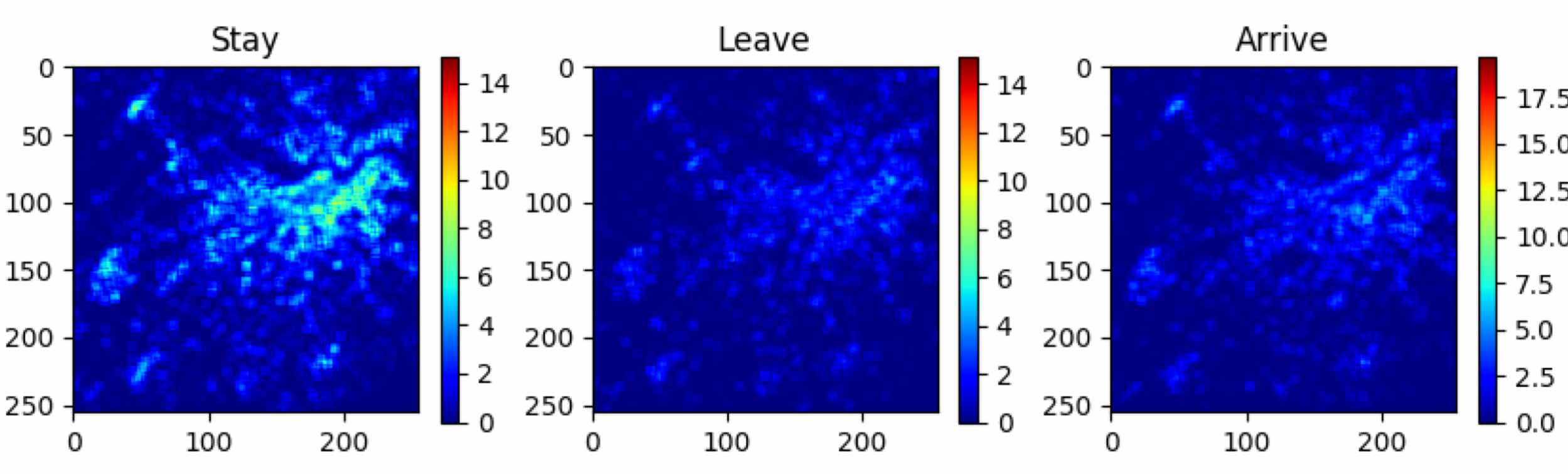}}\label{fig:sh_sleeping}
    \subfigure[Working]{
        \includegraphics[width=0.46\textwidth]{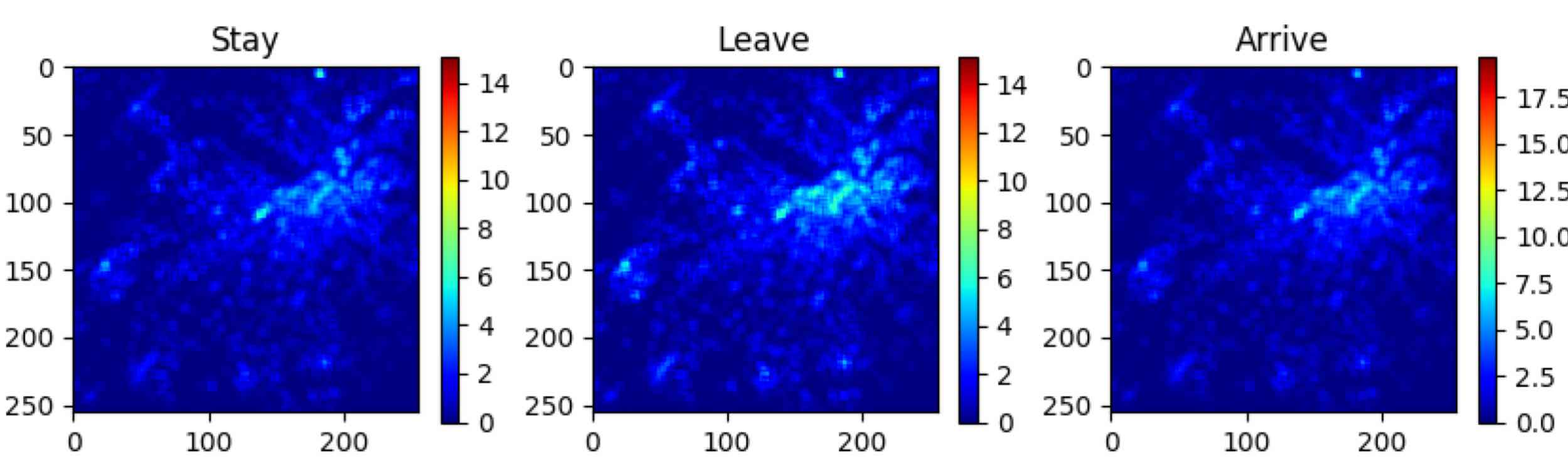}}\label{fig:sh_working}
  \caption{City images for Shanghai. We show the spatial distribution of the three original mobility features for four city states whose physical meaning are go-to-work rush, off-work rush, sleeping and working.}\label{fig:cityImageSH}
\end{figure}


\subsubsection{City Images for States}

To further explain the states obtained through hierarchical clustering, we show the spatial distribution of the three original mobility features for different time slots and compare their difference. 
Limited by space, we only compare \textbf{Morning rush}, \textbf{Afternoon rush}, \textbf{Sleep 1}, \textbf{Work 1} states in Shanghai, whose physical meanings are go-to-work rush, off-work rush, sleeping, working as shown Fig. \ref{fig:cityImageSH}. The heatmap is colored with the relative density.

1) In Shanghai, Compared with working state, people's staying is distributed more uniformly with low arriving and leaving in sleeping state. However, for working state, people are staying in some specific area with higher arriving and leaving than sleeping state. The reasonable explanation is that people are staying at home and the living area in the city is distributed more uniformly than office areas. 

2) As for go-to-work rush and off-work rush, the arriving-channel and leaving-channel have higher values than other states. The distribution of mobility in city address the downtown area and main road. These show that these two states are much about traffic. Interestingly, staying people in off-work rush are more than those in go-to-work rush. And this may due to that people usually have a uniform time to go to work, but do not have uniform off work time. Someone keep staying office while others are on the way home.
We also find that the arriving-channel and leaving-channel in go-to-work rush is similar to the leaving-channel and arriving-channel of off-work rush. This implies that off-work rush is the opposite process of go-to-work rush. 

\subsection{Long-term periodicity} 
We directly observe the long-term periodicity of urban dynamics from the full time mapping and 24-hour mapping visualization. 

From \ref{fig:bj_clustering}(a) and Fig. \ref{fig:sh_clustering}(a), we observe that there are several kinds of dynamics daily repeating, which well correspond to the weekday and weekend periodicity. Specifically, for Beijing, Qingming Festival and May Day also present different dynamics.

From \ref{fig:bj_clustering}(b) and Fig. \ref{fig:sh_clustering}(b), after dividing days into weekends, weekdays, Qingming Festival and May Day, the difference of different kinds of days could be observed. Besides, the same kind of days present  similar dynamics.

We conclude that the long-term periodicity of urban dynamics are caused by the periodicity of weekdays, weekends and festival holidays. The deeper reason is that residents tend to behave similarly in the same kind of days, while behave differently in different kinds of days. Readers could refer to our interpretation of city states to compare residents' different behaviors in different dynamics.

Another finding is that, urban dynamics are highly repeating not only in days, but also in hours. Hours-regularity are investigated by motif analysis in the next section.

\subsection{Short-term regularity}

\begin{figure*}[ht]
\centering
    \subfigure[Sleeping motif]{
        \includegraphics[width=0.25\textwidth]{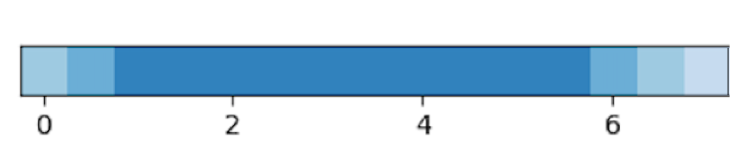}}\label{fig:sleep_motif}
    \subfigure[Weekday motif]{
        \includegraphics[width=0.8\textwidth]{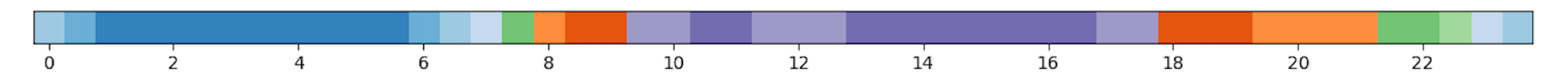}}\label{fig:weekday_motif}
\caption{Two motif examples. The colors in the motifs correspond to the city states in Fig. \ref{fig:bj_clustering}. The motif which covers 0 am to 7 am in each day is shown in (a), corresponding to residents' regular sleeping behavior. (b) is the motif which covers 0 am to 24 am in each weekday, corresponding to residents' regular behavior for sleeping(0-8 am), commuting(8-9 am, 18-21 am), working(9-18 am), relaxing(21-24 am) in a weekday. }
\label{fig:motif_example}
\end{figure*}

\begin{figure*}[ht]
\centering
{\includegraphics[width=1\textwidth]{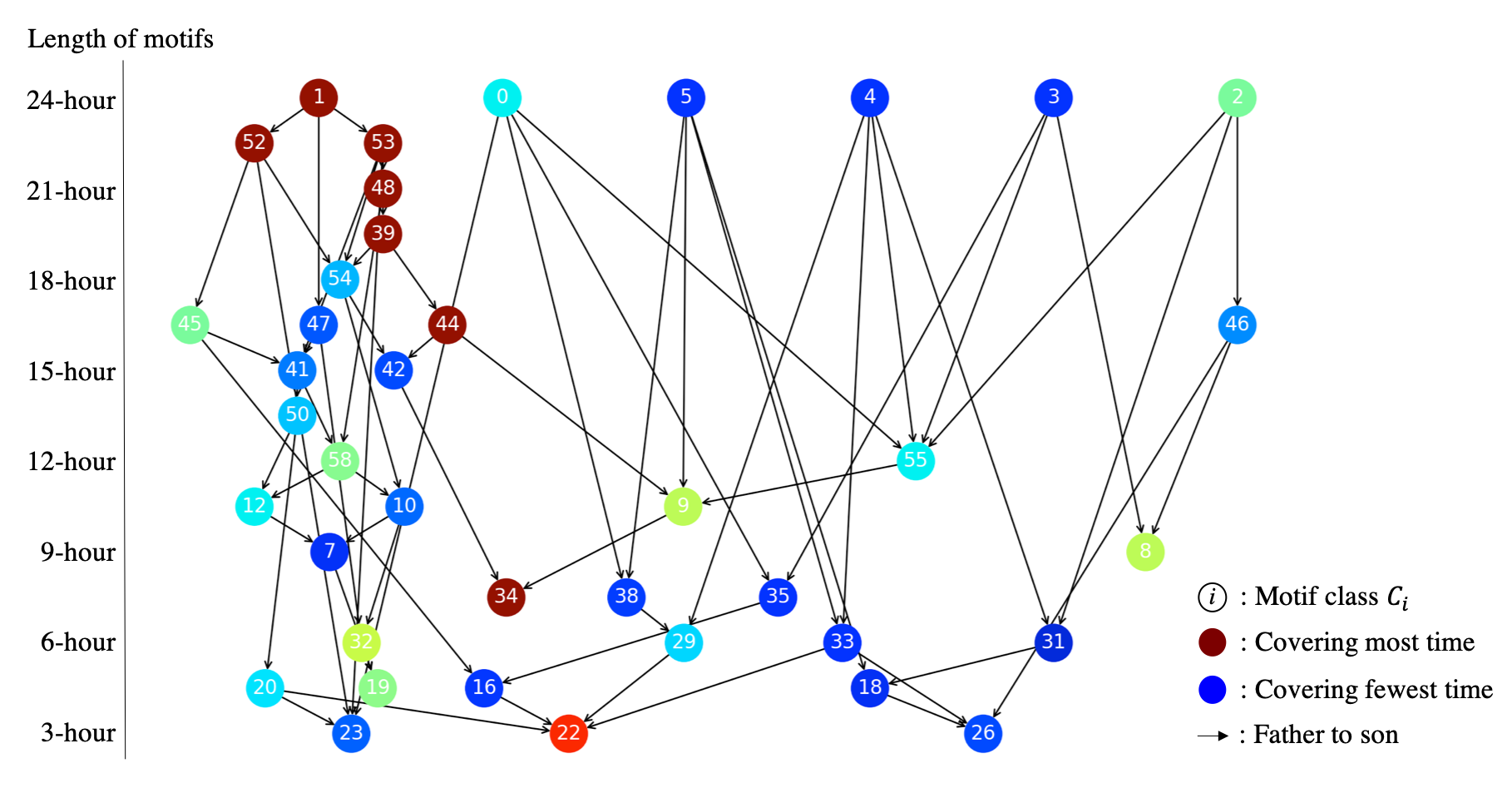}}
\caption{The motif family graph. Each node represents a motif class. Arrows are from father motifs to son motifs. The color of each node presents the total time this motif class covers. Nodes are arranged in height by the length of their representing motif classes.}
\vspace{-0.2cm}
\label{fig:motif_family}
\end{figure*}

As Shanghai dataset covers only 5 days and Beijing dataset covers 30 days, the regularity of urban dynamics in Beijing is more general and convincing. Thus, we only implement motif analysis on Beijing dataset.

\subsubsection{Implementation details}
We set $l_w=6,s_w = 2, \sigma_w =1$. To investigate short-term regularities, we limit each motif within a day and set a frequency threshold $f_{threshold}=3$, that only motifs with more than $f_{threshold}$ will be counted. An exception is for the motifs whose time lengths are 24 hours (we set $f_{threshold} = 1$ for them), for we wish to know all the kinds of daily dynamic the city will experience. In DBSCAN clustering, we set min samples to be 2 and eps to be 0.25 multiply the length of motifs (with a max eps threshold $eps_{max}=8$). Also, we ignore some nodes with out-degree and in-degree equal to 1 when plotting the motif family graph. These parameters and operations are for a clean but meaningful motif visualization. Parameters are chosen from experiments. We infer the corresponding resident behavior for each motif by analyzing how this motif is composed with different city states and when this motif appears.

\subsubsection{Motifs}

After employing our motif analysis method on the detected city state series, 59 motifs of different lengths are identified.
Two motif examples are shown in Fig.  \ref{fig:motif_example}. We refer two motif examples as \textbf{sleeping motif} and \textbf{weekday motif}, respectively. For the sleeping motif, it contains a sequence of Sleep 3 - Sleep 2 - Sleep 1 - Sleep 2 - Sleep 3 - Home. We consider it correspond to residents' sleeping behaviors. For the weekday motif, it starts from 0 am  and last to 24 am in weekdays, with a state transform from sleeping states, to rush states, to working states, to relaxing states and finally back to sleeping states. This motif well corresponds to residents' behaviors within a weekday and keep repeating in all weekdays.

\subsubsection{Motif relationships}

\begin{figure*}[ht]
\centering
    \subfigure[Holiday motif family]{
        \includegraphics[width=0.8\textwidth]{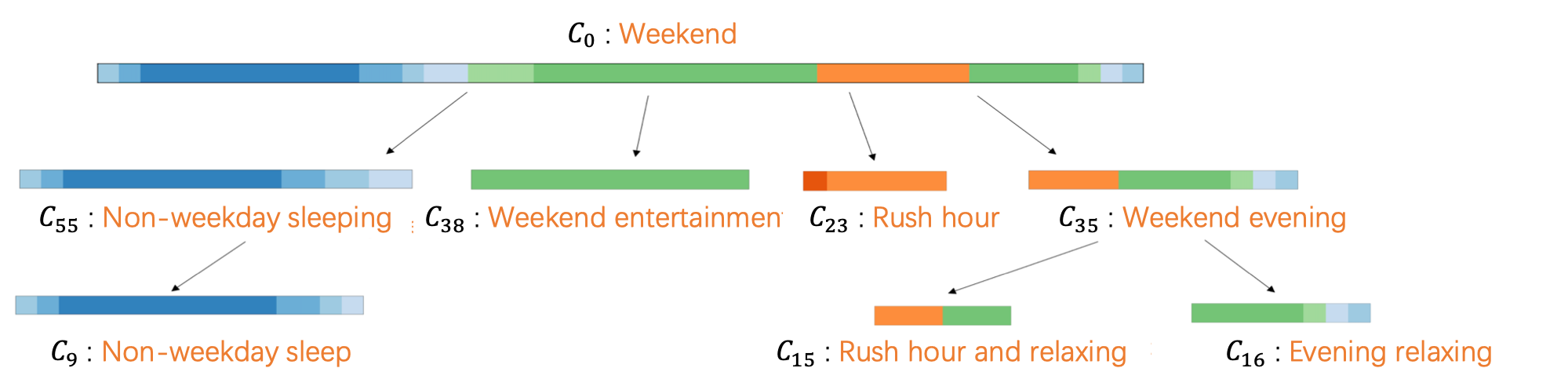}}
    \subfigure[Weekday motif family]{
        \includegraphics[width=0.8\textwidth]{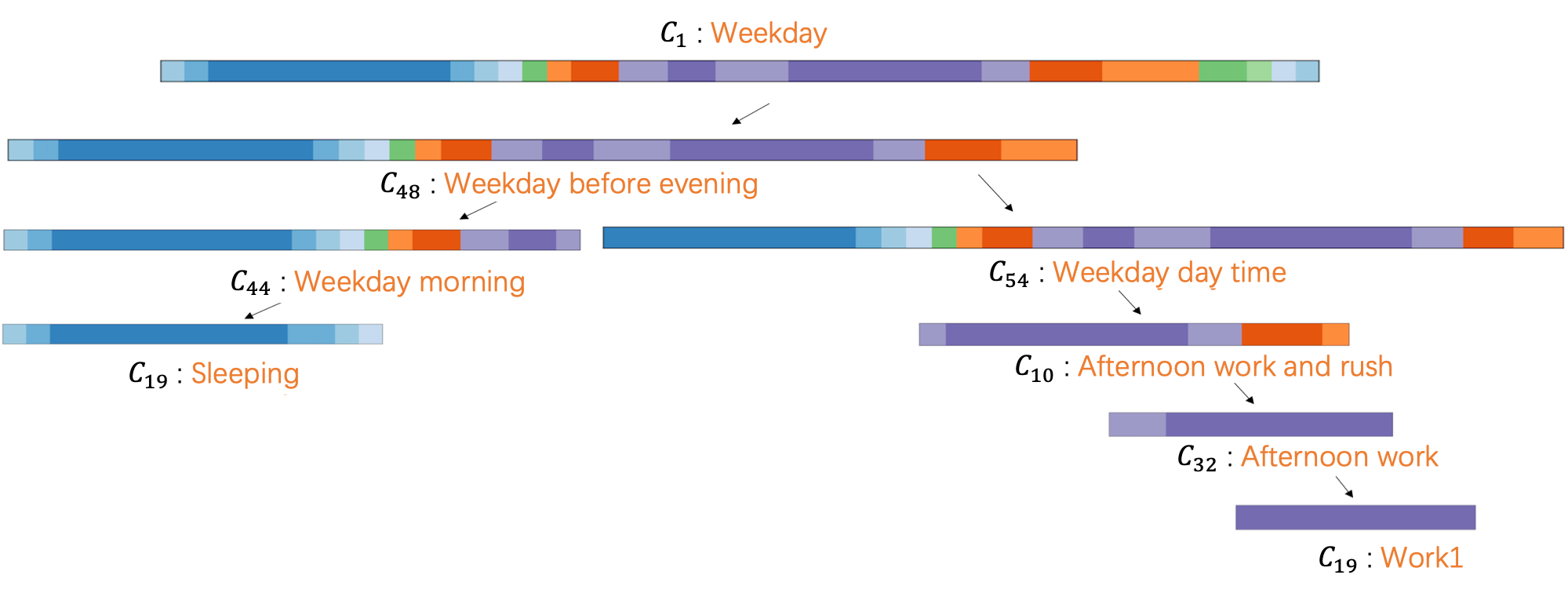}}
    \subfigure[Weekend motif family]{
        \includegraphics[width=0.8\textwidth]{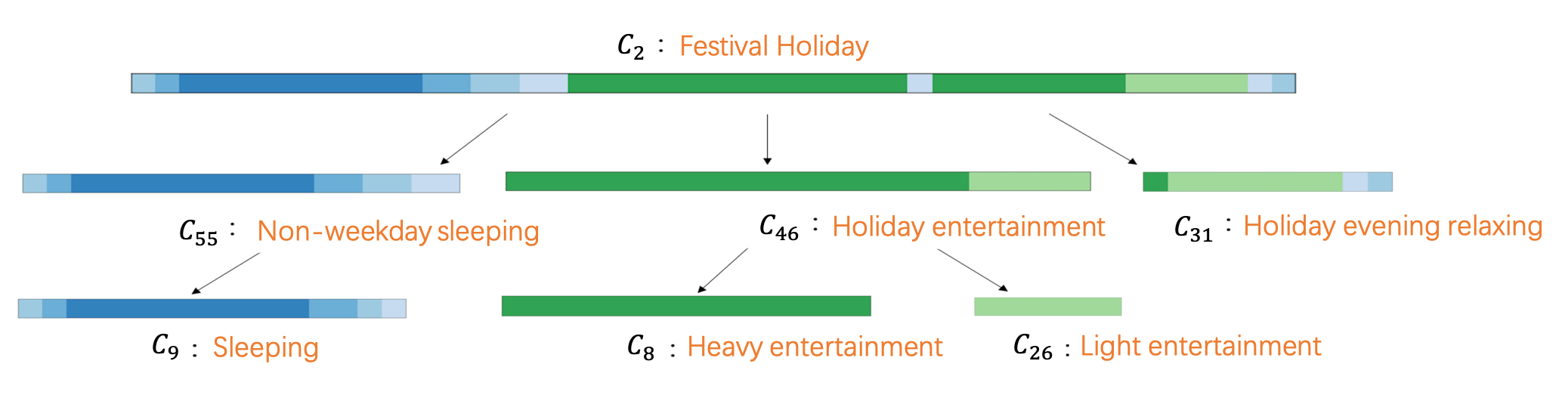}}
\caption{Visualization of motif families for $C_0$, $C_1$ and $C_2$. We use orange text to mark the corresponding residents' behaviors for each motif class. Due to the limit of space, we don't visualize all the motif classes in the families.}
\label{fig:familyexample}
\end{figure*}

Note that, the sleeping motif is a sub-sequence of the weekday motif, which means the regular sleeping behavior is a component of the regular weekday behavior. This kind of relationship also exist between other motifs. To further investigate the relationship of motifs, we plot a directed graph to present the hierarchy motif family, shown in Fig.  \ref{fig:motif_family}. In this graph, every node represent a motif class $C_i$. The inclusion relationship of motifs are represented by an directional edge, from the father motif to its son motif. For example, the sleeping motif $C_9$ is a component of the weekday motif $C_1$. Thus from the node with mark 1, go along with the arrow, pass the son and some grandsons, we can reach the node with mark 9.
In the figure, we also divide motifs into different layers according to their lengths. The top layer motifs have longest length, which is 24 hours time. While the lowest layer motifs have the shortest length, which is determined by the window length.

\begin{figure*}[ht]
\centerline{\includegraphics[width=0.8\textwidth]{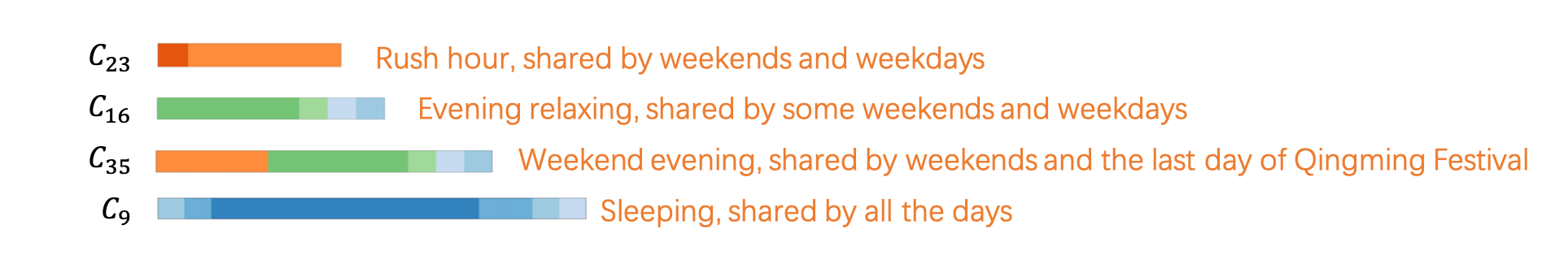}}
\caption{Short-term motifs shared by different top motifs. Each shared motif corresponds to a basic short-term regular behavior of residents.}
\vspace{-0.2cm}
\label{fig:shared_motif}
\end{figure*}

From Fig.  \ref{fig:motif_family}, we could observe 6 top motif classes, which means at the level of 24 hours, the city experiences 6 kinds of states transforms, and further indicates residents have 6 kinds of regular 24-hour behaviors. Among these 6 motif classes, $C_0$ occurs only in weekends, and corresponds to residents regular behaviors in weekends. $C_1$ occurs only in weekdays, and includes a working time motif. Thus we refer $C_1$ as weekday motif, which represents residents' behaviors in weekdays. We denote $C_2$ as holiday motif for the same reason. $C_3$, $C_4$, $C_5$ occurs only once in the 30 days of Beijing dataset. 

We could observe that $C_0$, $C_1$, $C_2$ has lush families which contain a lot of sons and grandsons. This means these long-term regular behaviors is composed with many short-term regular behaviors. To dig deeper into it, we visualize how a long-term motif is composed with short-term motifs and analyze how this related to people's behaviors. The visualization is shown in Fig.  \ref{fig:familyexample}.

Interestingly, some short-term motifs are shared by many different long-term motifs. We visualize the shared motifs in Fig. \ref{fig:shared_motif}.
We speculate this kind of short-term motifs correspond to people's basic behaviors. Sleeping, as a very basic behaviors, are shared by all days. Though different days may have different form of sleeping behaviors (see $C_{57}$ and $C_{55}$), a basic sleeping is common (see $C_9$ and $C_6$). Another interesting finding is $C_{35}$, that the last evening of Qingming festival shares the same $C_{35}$ with weekends evenings. (see the analysis below)

\subsection{Special dynamic Patterns}
By observing dynamics in Fig. \ref{fig:bj_clustering} and Fig. \ref{fig:sh_clustering} and comparing different motifs, we find some interesting dynamic patterns. Some of our finding well match people's intuition while some give surprises.

\textbf{Weekends vs Holidays}: 
Two holidays are detected through our method, i.e., Qingming Festival and May Day. People have intuition that weekends and holidays are different, but wonder why and how.
As shown $C_0$ and $C_2$, in weekends, Relax 3 covers very morning time and Relax 2 covers other day time and some evening time. However in holidays like Qingming Festival and May day, Relax 1 covers almost all the time. Relax 3 covers very morning time and almost all the evening time. This shows that people's movements pattern are similar in weekends' and holidays' mornings and evenings, while differ in their day-time. we conclude that in holidays' day-time, people tend to travel far away from the city, while in mornings and evenings,  people haven't set off or have backed the city, following the same pattern as weekends.

\textbf{Last evening of holidays}: 
We usually have a sense that on the last evening of holidays, our pace of life back to normal.
Interestingly, as shown in Fig. \ref{fig:shared_motif}, we find that $C_{35}$ is shared by both weekends evenings and the evening night of Qingming Festival. This indicates that in the last evening of Qingming Festival, the city's dynamic back to weekends patterns, where a Rush 2 state appears first, then followed Relax 2 and Relax 3. It matches with our intuition that people come back city in the last day of holiday, causing a traffic jam, then most people get home while some people still hang out.
Note that our data only covers the first two days of May Day, so this motif doesn't appear in May Day.

\textbf{Symmetric night}:
We find that the sleeping behaviors of residents are more symmetric than expected. This pattern is for all the days, regardless weekdays or not. As shown in $C_9$, city's states in night are : Home - Sleep 3 - Sleep 2 - Sleep 1 - Sleep 2 - Sleep 3 - Home.  Though this comes from people's movement patterns, but well matched people's sleeping habits. The government can properly arrange resources like illumination and construction according to this night dynamics.

\textbf{Unexpected peace in afternoons}:
We find Home state surprisingly appears in two holiday afternoons and one weekend afternoon. This suggests people's slight movement, which means at these moments, the city is as 'quite' and 'peaceful' as about-to-sleep hours.

\subsection{Validation with App Usage}
\begin{figure}[htb]
\centerline{\includegraphics[width=0.4\textwidth]{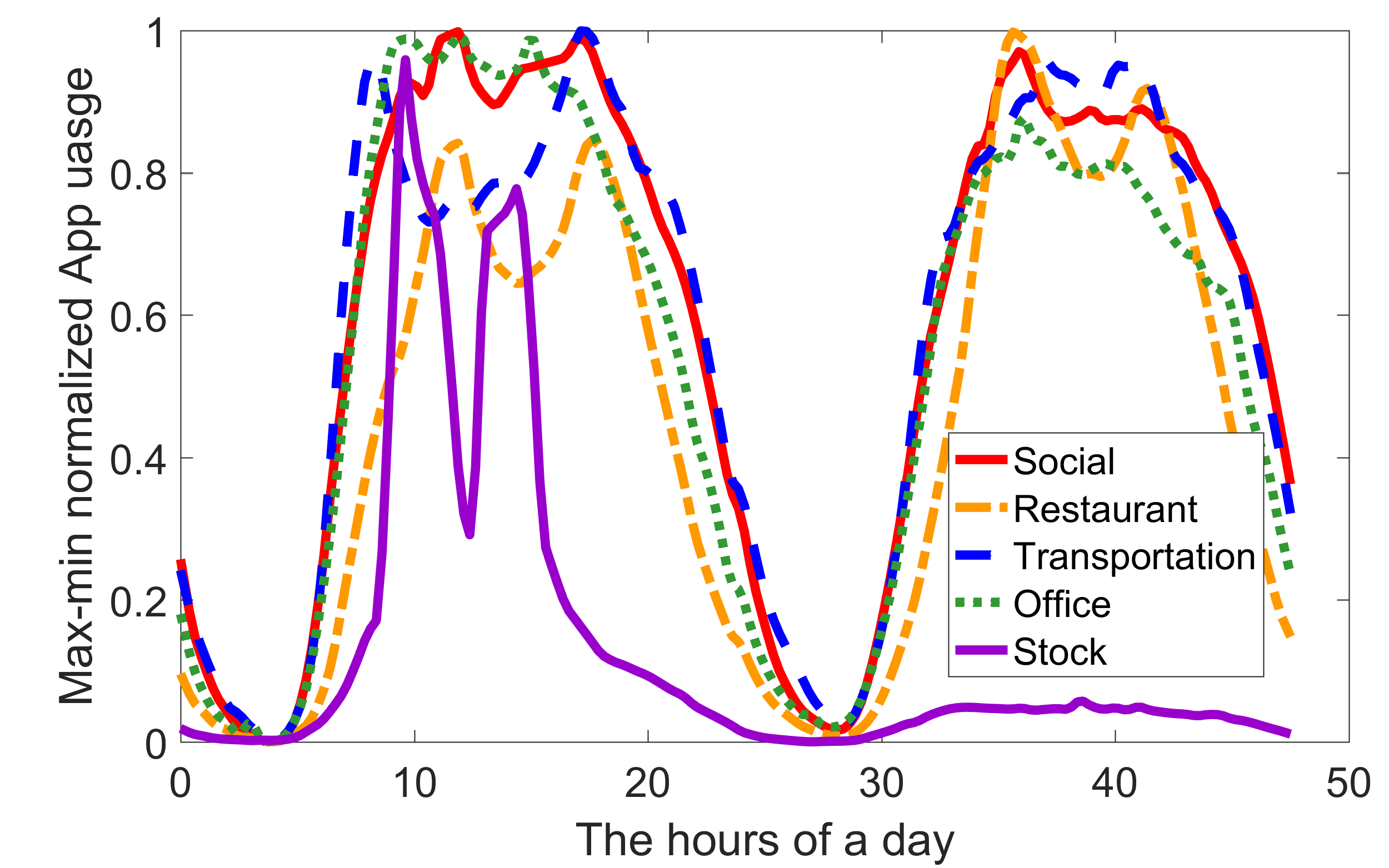}}
\caption{Normalized App usage in Shanghai, where the first 24 hours represent weekday, while the last ones represent weekend.}
\label{fig:app}
\end{figure}

In order to validate our explanations of urban dynamics through App usage, we first analyze the normalized App usage at different time slots. 
To make it clear, we show the normalized curve of some Apps in Fig. \ref{fig:app}. From these figures, we can observe that the usage of different Apps are various in hour and day. For example, Stock Apps are used most frequently during 9:00-11:00 and 13:00-15:00, which is Stock market time. 
Thus, App usage could be utilized to validate the states we identify, and further to explain the urban dynamics  from these states.

\begin{table*}[ht]
\begin{center}
\scalebox{0.9}{
\begin{tabular}{c|ccccccccc}
  \hline
Usage & \textbf{Morning rush} & \textbf{Afternoon rush} & \textbf{Sat morning} & \textbf{Relax 1}  & \textbf{Relax 2} & \textbf{Home} & \textbf{Work 1} & \textbf{Work 2} & \textbf{Sleep}\\
    \hline
\textbf{Social} & 0.326  & 0.418***  & 0.317  & 0.351  & 0.338***  & 0.22***  & 0.409  & 0.4  & 0.073  \\
\textbf{Video} & 0.342  & 0.412  & 0.336**  & 0.361***  & 0.36*  & 0.238*  & 0.376  & 0.367  & 0.091*  \\
\textbf{Music} & 0.446*  & 0.436**  & 0.302  & 0.316  & 0.309  & 0.217  & 0.344  & 0.391  & 0.07  \\
\textbf{Reading} & 0.441**  & 0.391  & 0.321  & 0.318  & 0.291  & 0.207  & 0.393  & 0.402  & 0.068  \\
\textbf{Game} & 0.398***  & 0.411  & 0.323  & 0.34  & 0.323  & 0.229**  & 0.379  & 0.381  & 0.087**  \\
\textbf{Shopping} & 0.338  & 0.413  & 0.31  & 0.353  & 0.308  & 0.189  & 0.427***  & 0.423***  & 0.059  \\
\textbf{Restaurant} & 0.226  & 0.461*  & 0.332***  & 0.439*  & 0.355**  & 0.143  & 0.398  & 0.357  & 0.041  \\
\textbf{Transportation} & 0.354  & 0.402  & 0.369*  & 0.374**  & 0.304  & 0.208  & 0.386  & 0.385  & 0.076  \\
\textbf{office} & 0.356  & 0.392  & 0.316  & 0.334  & 0.306  & 0.206  & 0.429**  & 0.424**  & 0.086***  \\
\textbf{stocks} & 0.192  & 0.195  & 0.074  & 0.062  & 0.092  & 0.058  & 0.815*  & 0.489*  & 0.016  \\
     \hline
\end{tabular}
}
\end{center}
\caption{The TF-IDF results for App usage, where * means the most frequently used APP in each state, while ** and *** means the second and the third frequently used APP in each state, respectively.}\label{tab:app}
\end{table*}

Considering that the numbers of Apps in each App category are different, we can not compare the absolute usage count in the same state directly. In order to address this problem, we use TF-IDF statistic to analyze the relationship between App usage and city states \cite{Paik2013}. We denote $U$ as the absolute usage count of each App, where $U_{i,j}$ means the usage of  $i$-$th$ App under $j$-$th$ state. Thus, the transformed App usage $U'$ can be calculated as follows,
\begin{equation}
U'_{i,j}=\frac{U_{i,j}}{\sum\limits_{j}{U_{i,j}}} \times \log\frac{\sum\limits_{i}{U_{i,j}}}{U_{i,j}}.
\end{equation}
The result is shown in Table \ref{tab:app},where we can observe that:\\
1) In Sleep state, the usage of all Apps are lowest. \\
2) In Work states, including Work 1 and Work 2, the usage of Stock and Office are highest. \\
3) In Rush states, the usage of Transportation Apps is high in Morning rush, Afternoon rush, and highest in Saturday morning. Interestingly, in Morning rush and Afternoon rush, the usage of Music and Restaurant is highest. \\
4) In Relax states, including Relax 1 and Relax 2, the usage of Restaurant, Video, Transportation, Social Apps are high. Specifically, in Relax 1 state where some people tend to travel far in weekends, the usage of Transportation Apps is higher than that in Relax 2 state where people get fewer movement.\\
5) In Home state, the usage of all Apps is low and the usage of Video and Game are highest among them. People tend to stay home, rest and relax. 

These observations and conclusions support our interpretation for the identified city states, and further demonstrate that urban dynamics could be revealed from human mobility behaviors.

\section{Discussion}
Through UrbanRhythm, we acquire knowledge about the basic city states, the long-term periodicity and the short-term regularity of urban dynamics. We further analyze how each correspond to residents' behaviors. This  provide comprehensive knowledge and interpretable support for city governing. Governors could give different resource schedules according to different dynamics at different time.
Besides, UrbanRhythm provides a tool for social research. For example, social researchers could  compare urban dynamics between different cities to study how regional life style is differ accorss cities.

After the revealing of urban dynamics, the next question is what factor would affect urban dynamics and residents' behaviors. 
Possible answers include short-term factors such as an emergency event or special weather, and long-term factors like seasons or economy. These need further investigations.

\section{Related Work} \label{sec:Relatedwork}
\textbf{Urban dynamics modeling:}
Forrester first summarized the previous researches about modeling bits and pieces of urban areas as urban dynamics models in \cite{urbandy1}. \cite{kadanoff1971from,belkin1972urban}  proved and extended the model proposed by Forrester. In addition, Batty et al. \cite{BATTY1999205} utilized fine-grained cellular automata to model urban activities, which can be adapted to simulate urban development over very different time period. 
In recent years, \cite{cranshaw2012livehoods} detected city areas depicting a snap-shot of activity patterns of its people. With more attention to temporal dimension, \cite{kling2012city} used a Topic model to characterize urban dynamics; \cite{Zhang2013} used the geo-tagged social data to analyzed urban dynamics; \cite{fan2014cityspectrum} modeled city dynamics in a basic life pattern space. We also reveal urban dynamics from the view of temporal dimension. Different from previous works, we divide the city into different hierarchical states and characterize urban dynamics as the transform of city states. Moreover, we consider the spatial distribution of human mobility in the city as a factor influencing urban dynamic and use an image processing method to capture such distribution patterns.

\textbf{Mobility pattern revealing:}
Revealing the hidden pattern in mobility data becomes popular these years \cite{ratti2006mobile, reades2007cellular, candia2008uncovering, blondel2015survey}.
From the view of individuals, \cite{qin2018spatio,thuillier2018clustering,noulas2012tale} revealed the pattern of people's behaviors. From the view of regions,  \cite{andrienko2013scalable} explored significant places; \cite{yuan2012discovering, zhao2015automatic} predicted the function of regions; \cite{fan2014cityspectrum} used a non-negative tensor factorization approach to decompose human mobility into variations among regions and times; \cite{xia2019revealing} revealed the daily activity pattern of specific regions. From the view of events, \cite{6876004} detected special event by analyzing spatio-temporal data; \cite{calabrese2010geography} analyzed cell-phone mobility and the relationship between events and attendees. 
To best of our knowledge, we are first to use mobility data to understand urban dynamics from the view of the whole city. Our analyzing target is not a single region, but the whole city composed with numerous regions. Thus we use image processing method Saak to capture the spatial distribution pattern of human mobility. Our analysis of App usage gives more interpretation to our results.

\textbf{Image transformation and its application:} 
In this paper, we use Saak transform \cite{saak} to extract the spatial distribution pattern of mobility for city images. Saak transform is a spatial-spectral transform like the discrete cosine transform \cite{DCT} and the Wavelet transform \cite{wavelet}. There have been a lot of applications for these transforms, like image coding \cite{imagecoding}, image compression \cite{watson1994image}, face recognition \cite{hafed2001face}, etc. 
To best of our knowledge, we are the first to apply image transformation  in urban dynamics detection. 
There are also deep learning methods for image transform, i.e., unsupervised feature extraction \cite{dosovitskiy2014discriminative,joint,gan}. However, they are hard to train and require a large number of training samples, making it not realistic in our problem.

\section{Conclusion}

In this paper, we propose a novel system UrbanRhythm to reveal urban dynamics hidden in mobility data. We divide the city into different time slots, calculate the mobility feature in each time slot and classify those time slots into hierarchical city states. Then, we characterize the urban dynamics as the transform of city states along time axis. We further observe long-term periodicity from the visualization of urban dynamic and use motif analysis to investigate the short-term regularity. Besides, the detected city states and regularity of urban dynamics are interpreted, which pave the way for more applications, such as traffic monitoring and resource scheduling.
Extensive experiments on two real-life datesets of different cities demonstrate the efficiency of our method. An App usage analysis also is used to validate our interpretation of city states. 
Our work opens a new perspective to investigate urban dynamics and to reveal the patterns in mobility data.

\ifCLASSOPTIONcompsoc
  \section*{Acknowledgments}
\else
  \section*{Acknowledgment}
\fi
This work was supported in part by The National Key Research and Development Program of China under grant SQ2018YFB180012, the National Nature Science Foundation of China under 61971267, 61972223, 61861136003, and 61621091, Beijing Natural Science Foundation under L182038, Beijing National Research Center for Information Science and Technology under 20031887521, and research fund of Tsinghua University - Tencent Joint Laboratory for Internet Innovation Technology.

\bibliography{cite}
\bibliographystyle{abbrv}


\begin{IEEEbiography}[{\includegraphics[width=1.2in,height=1.4in,clip,keepaspectratio]{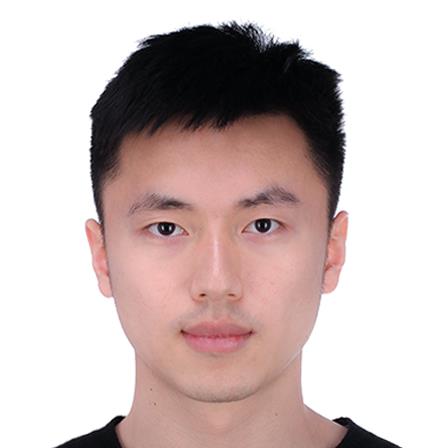}}]{Sirui Song} is an undergraduate student in department of Electronic Engineering, Tsinghua University, Beijing, China. His work mainly focuses on spatio-temporal data mining. \end{IEEEbiography}

\begin{IEEEbiography}[{\includegraphics[width=1.0in,height=1.4in,clip,keepaspectratio]{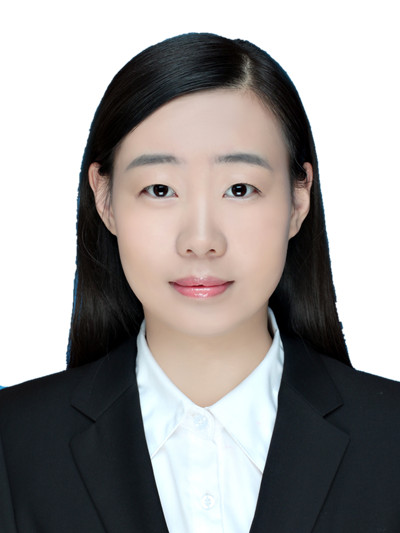}}]{Tong Xia} received the B.S. degree in electrical engineering from School of Electrical Information, Wuhan University, Wuhan, China, in 2017. At present, she is studying for the M.S. degree in big data from Department of Electronic Engineering, Tsinghua University, Beijing, China. Her research interests include human mobility, mobile big data mining, user behavior modelling and urban computing. \end{IEEEbiography}

\begin{IEEEbiography}[{\includegraphics[width=1.0in,height=1.4in,clip,keepaspectratio]{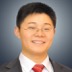}}]{Yong Li}(M'09-SM'16) received the B.S. degree in electronics and information engineering from Huazhong University of Science and Technology, Wuhan, China, in 2007 and the Ph.D. degree in electronic engineering from Tsinghua University, Beijing, China, in 2012. He is currently a Faculty Member of the Department of Electronic Engineering, Tsinghua University.
Dr. Li has served as General Chair, TPC Chair, SPC/TPC Member for several international workshops and conferences, and he is on the editorial board of two IEEE journals. His papers have total citations more than 6900. Among them, ten are ESI Highly Cited Papers in Computer Science, and four receive conference Best Paper (run-up) Awards. He received IEEE 2016 ComSoc Asia-Pacific Outstanding Young Researchers, Young Talent Program of China Association for Science and Technology, and the National Youth Talent Support Program.
\end{IEEEbiography}

\begin{IEEEbiography}[{\includegraphics[width=1.0in,height=1.4in,clip,keepaspectratio]{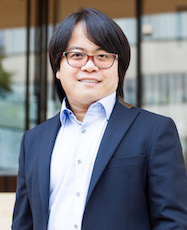}}]{Pan Hui} (F’17) received the B.Eng. and M.Phil. degrees, both from the Department of Electrical and Electronic Engineering, University of Hong Kong, Hong Kong, and the Ph.D. degree in computer laboratory from the University of Cambridge, Cambridge, U.K. He is currently a Faculty Member with the Department of Computer Science and Engineering, Hong Kong University of Science and Technology, Hong Kong, where he directs the System and Media Lab. He also serves as a Distinguished Scientist with Telekom Innovation Laboratories (T-labs) Germany and an Adjunct Professor of social computing and networking, Aalto University, Finland. Before returning to Hong Kong, he has spent several years in T-labs and Intel Research Cambridge. He has published more than 100 research papers and has several granted and pending European patents. He has founded and chaired several IEEE/ACM conferences/workshops, and served on the technical program committee of numerous international conferences and workshops including IEEE Infocom, SECON, MASS, Globecom, WCNC, and ITC.
\end{IEEEbiography}

\begin{IEEEbiography}[{\includegraphics[width=1.0in,height=1.4in,clip,keepaspectratio]{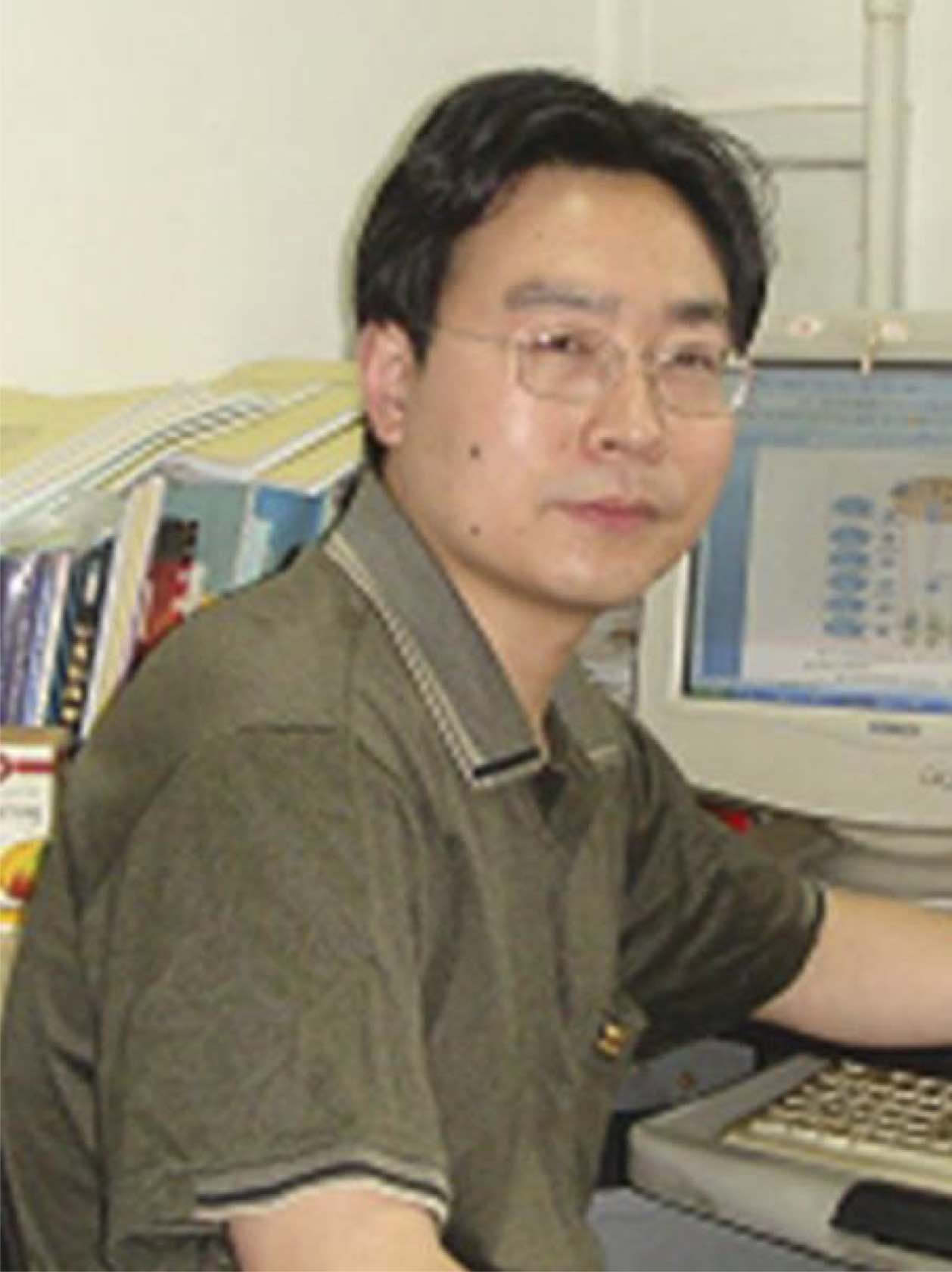}}]{Depeng Jin} (M'2009) received his B.S. and Ph.D. degrees from Tsinghua University, Beijing, China, in 1995 and 1999 respectively both in electronics engineering. Now he is an associate professor at Tsinghua University and vice chair of Department of Electronic Engineering. Dr. Jin was awarded National Scientific and Technological Innovation Prize (Second Class) in 2002. His research fields include telecommunications, high-speed networks, ASIC design and future internet architecture.
\end{IEEEbiography}

\end{document}